\documentclass[acmsmall]{acmart}
\usepackage{pdflscape}
\usepackage{mdframed}

\usepackage{changepage} 

\usepackage{array}
\newcolumntype{P}[1]{>{\centering\arraybackslash}p{#1}}

\AtBeginDocument{%
  \providecommand\BibTeX{{%
    \normalfont B\kern-0.5em{\scshape i\kern-0.25em b}\kern-0.8em\TeX}}}

\setcopyright{rightsretained}
\acmJournal{PACMHCI}
\acmYear{2024} \acmVolume{8} \acmNumber{CSCW2} \acmArticle{466} \acmMonth{11}\acmDOI{10.1145/3687005}

\received{July 2023}
\received[revised]{January 2024}
\received[accepted]{March 2024}

\begin{document}

\title[Insights from a data crowdsourcing experiment]{Insights from an experiment crowdsourcing data from thousands of US Amazon users: The importance of transparency, money, and data use}


\author{Alex Berke}
\email{aberke@mit.edu}
\orcid{0000-0001-5996-0557}
\affiliation{%
  \institution{MIT Media Lab}
  \country{USA}
}

\author{Robert Mahari}
\email{rmahari@mit.edu}
\orcid{0000-0003-2372-2746}
\affiliation{%
  \institution{MIT Media Lab}
  \country{USA}
}

\affiliation{%
  \institution{Harvard Law School}
  \country{USA}
}

\author{Sandy Pentland}
\email{pentland@mit.edu}
\orcid{0000-0002-8053-9983}
\affiliation{%
  \institution{MIT Media Lab}
  \country{USA}
}
\affiliation{%
  \institution{Stanford HAI}
  \country{USA}
}

\author{Kent Larson}
\email{kll@mit.edu}
\orcid{0000-0003-4581-7500}
\affiliation{%
  \institution{MIT Media Lab}
  \country{USA}
}

\author{Dana Calacci}
\email{dcalacci@psu.edu}
\orcid{0000-0002-9552-1137}
\affiliation{
    \institution{Penn State University}
    \country{USA}
    }
\affiliation{
    \institution{MIT Media Lab}
    \country{USA}}



\renewcommand{\shortauthors}{Berke, Mahari, Pentland, Larson, \& Calacci}

\begin{abstract}
Data generated by users on digital platforms are a crucial resource for advocates and researchers interested in uncovering digital inequities, auditing algorithms, and understanding human behavior. Yet data access is often restricted. How can researchers both effectively and ethically collect user data? This paper shares an innovative approach to crowdsourcing user data to collect otherwise inaccessible Amazon purchase histories, spanning 5 years, from more than 5,000 U.S. users. We developed a data collection tool that prioritizes participant consent and includes an experimental study design. The design allows us to study multiple important aspects of privacy perception and user data sharing behavior, including how socio-demographics, monetary incentives and transparency can impact share rates. Experiment results (N=6,325) reveal both monetary incentives and transparency can significantly increase data sharing. Age, race, education, and gender also played a role, where female and less-educated participants were more likely to share. Our study design enables a unique empirical evaluation of the “privacy paradox”, where users claim to value their privacy more than they do in practice. We set up both real and hypothetical data sharing scenarios and find measurable similarities and differences in share rates across these contexts. For example, increasing monetary incentives had a 6 times higher impact on share rates in real scenarios. In addition, we study participants' opinions on how data should be used by various third parties, again finding that gender, age, education, and race have a significant impact. Notably, the majority of participants disapproved of government agencies using purchase data yet the majority approved of use by researchers. Overall, our findings highlight the critical role that transparency, incentive design, and user demographics play in ethical data collection practices, and provide guidance for future researchers seeking to crowdsource user generated data.
\end{abstract}



\maketitle

\section{Introduction}

The data that users generate on digital platforms is a critical resource for users interested in understanding their own data, researchers who audit algorithms or study human behavior, and advocates interested in holding digital platforms to account. Recent important audits of gig economy pay algorithms~\cite{calacci2022bargaining}, online recommendation algorithms~\cite{ricksDoesThisButton2022}, and the online advertising ecosystem~\cite{berke2022privacy} have all relied on detailed user-generated data on user behavior. 

However, those interested in similar work that "studies up"~\cite{barabas2020studying} to examine how platforms impact users and society at large face major data access challenges. Companies such as Twitter charge increasingly large fees for API use, limiting researcher access~\cite{jingnan2023twitter}. Other platforms that provide access often place such strict limits on data use that they resemble a form of academic gatekeeping, allowing certain researchers and projects through while blocking others~\cite{lurie2023comparin}.

In light of these access issues, crowdsourced data collection has emerged as a strategy that enables independent access to user data. Crowdsourcing data, where users are compensated for or volunteer their data to a project, lies in stark contrast to using APIs developed by companies or datasets already available for purchase or research in several ways. Crowdsourcing user data has the potential to engage participants in a more informed consent process, opens the possibility for more participatory research designs, and can help provide users themselves with access to their data. 

However, crowdsourcing has its own significant challenges. First, it depends on users willingly sharing their data. This makes studying how, and why, users might share their data with a project an important consideration. For example, those seeking a representative sample of users might risk over- or under-representing certain groups by offering cash incentives for data sharing. Designers of crowdsourcing efforts also lack guidance on what methods are most effective for recruitment and maximizing data sharing while maintaining ethical standards. 
Furthermore, there is no standard of conduct or set of guidelines for the collection, use, and governance of data that people contribute to research projects outside of IRB guidelines.

What incentives, designs, and framings impact users’ likelihood of sharing personal data already collected by platforms?
While recent studies highlight the importance of motivating users to donate personal data for research, and begin to study what factors impact their likelihood to do so, these studies have been limited to hypothetical thought experiments, where users are asked about hypothetical sharing scenarios~\cite{skatovaPsychologyPersonalData2019, pfiffner2023leveraging}. 
In contrast, this work studies these questions by putting users in real sharing scenarios. 

Additionally, many studies ignore an important ethical consideration about user data: how, and by whom, users would like their data to be used. Contextual understandings of privacy support~\cite{nissenbaum2020privacy} the idea that to ethically use data shared by users, how users would (and would not) like their data to be used must first be understood. Current data sharing norms within human behavioral research often ignore this question and treat the ethical uses of data as a question to be answered by investigators, not study subjects, especially with respect to data stripped of personally identifiable information. What limits would users place on the use of the data they generate and share if they had a choice?

In this paper, we shed light on these questions by presenting findings from an innovative approach to crowdsourcing a dataset of purchase histories from over 6,000 crowd workers who shop on Amazon. We first share our tool design and experimental approach as a design template for future researchers interested in crowdsourcing user data. Our design focuses on participant consent and features an opt-in approach to data sharing, where, after accessing their purchase history data, participants actively choose or decline to share the data.

Using the opt-in design, we test how different conditions and monetary incentives impact the likelihood that participants choose to share their data. We manipulate two main variables. First, data transparency—whether users are shown their data when prompted to share. Second, we offer different monetary incentives ranging from \$0.05 to \$0.50 in exchange for users’ data. This allows us to test the impact of cash incentives on data sharing rates. 
We also test how participants’ hypothetical behavior compares to those in a real-world data exchange. In each study arm, some participants choose to share, and some decline. We place participants who decline to share their data in a “hypothetical” scenario, and ask them if they would hypothetically share their data if, in the future, they were offered an increasing menu of cash incentives. This hypothetical setup allows us to test the prevalence and magnitude of the privacy paradox: the observation that people appear to value their privacy more than their behavior suggests~\cite{norberg2007privacy}.

Additionally, we collect socio-demographic data from participants and survey their opinions on how they believe their purchase data should be used, and by what parties. This allows us to test for significant differences in sharing by demographic groups, and to report on how potential crowdsourcing participants want their data to be used by third parties.

Our results suggest that data transparency---showing participants their data before asking them to share---can significantly increase the rate at which participants choose to share their data. This can have a comparable effect to a monetary bonus, potentially offering a less expensive and arguably more ethical approach to crowdsourcing data. We also add a nuanced perspective to the ongoing privacy literature by measuring similarities and differences in the hypothetical and real sharing scenarios. In both contexts, we find that higher monetary incentives predictably increase share rates, but where the impact of increased incentives is lower in the hypothetical scenarios. More broadly, this work represents one of the first major empirical studies of data sharing behavior and the privacy paradox.

The remainder of the paper is organized as follows. First, we discuss in more detail some of the main challenges inherent in independent, crowdsourced data collection, and review work related to data privacy, user behavior in data sharing, and data donation. Then, we present our survey method and the experimental design we used to measure the effectiveness of different incentive schemes and framings. In the next section, we describe our data collection process, including preprocessing steps and statistics about our participants. The next three sections discuss our main findings on demographics and data sharing behavior, the impact of monetary incentives on share rates, and finally, survey results illustrating our participants’ normative views on data use. The discussion section assimilates our findings and shares implications for future researchers interested in crowdsourcing their own datasets.

\section{Background \& Related Work}
\label{sec:background_related_work}

\subsection{Privacy perception and data valuation
}
\label{sec:privacy_perception_data_valuation}

As new ways to generate, collect, and share data have multiplied, a diverse literature has developed that attempts to define privacy and characterize how individuals perceive and value it. More broadly, scholars have characterized privacy as the right to be left alone, the right to control personal information, and as an abstract social good, among other definitions~\cite{nissenbaum2020privacy,solove2008understanding}.
However, a general consensus has emerged in both legal and privacy studies communities that privacy is fundamentally defined by the social context in which one’s information is being used, a theory coined as contextual integrity by Helen Nissenbaum~\cite{nissenbaum2020privacy,cofone2022privacy}. 

Individual expectations and contextual norms about information flows then define what data should be private; privacy is violated when an actor or service knows more about an individual than that individual expects~\cite{cofone2017privacy}. 
This means that, for example, users might expect a weather application to use their location data, but not expect a banking application or music streaming service to collect their location~\cite{sweeney2002k}. 
A 2016 study that surveyed over 1,900 people found that while mobile app companies often did not meet user expectations, context—an app’s main purpose—significantly mediated what data respondents deemed acceptable to collect~\cite{bowser2017accounting}. 
Effects such as these are consistent and large, and provide convincing evidence for contextual definitions of privacy.

Many studies use survey-based methodologies to measure people’s expectations about privacy~\cite{cottrill2015location,morey2015customer}.
However, these methods face serious limitations. Using surveys alone to measure privacy expectations risks the privacy paradox: people consistently disclose more personal information than they claim to be comfortable with~\cite{norberg2007privacy,glasgow2021survey,spiekermann2001privacy}. 
Although the direction and magnitude of the privacy paradox have seen debate in recent years, it is generally accepted that there is a disconnect between user behavior and stated values when it comes to data privacy~\cite{colnago2023there,solove2021myth}. 
Work consistent with the contextual integrity theory of privacy has demonstrated that individuals largely lack economic rationality with regard to privacy decisions~\cite{acquisti2005privacy}, 
and that the concerns people express about privacy do not correlate with the amount of information people share publicly online~\cite{acquisti2006imagined}.

Previous work attempts to measure the monetary value that individuals place on privacy or their data using both field experiments and survey-based research. 
One interesting finding is that people suffer from an endowment bias when considering the value of their privacy: 
people value privacy more when they have it than when they do not~\cite{acquisti2013privacy,tang2021chinese}. 
This has been studied in the behavioral economics literature by comparing users’ willingness to pay (WTP) for privacy (to not have their data shared) and the value they are willing to accept (WTA) for disclosing personal information~\cite{winegar2019much, athey2017digital}. 
Our experimental design focuses on WTA, the amount of money users will accept in exchange for sharing data that has already been collected from them by another party (Amazon).

While much of the previous work measuring data valuation uses hypothetical survey-based methods, this paper extends recent work that attempts to measure the real-world value that participants are willing to accept in exchange for their data~\cite{li2021valuing} by actually transacting with study participants. 
In contrast to the existing literature, our experimental design allows us to quantify the difference between the amount participants in a hypothetical scenario say they are willing to accept for their data versus the actual amount that participants accept in a real transaction.

One major difference between our study and past literature is that we explicitly examine the impact of data transparency on how much participants value their data and opt to share it. 
While a recent study examining participants’ WTP and WTA values for location data tracked through a custom app showed participants the data in question, it did not measure the impact of data transparency on user behaviors~\cite{schmitt2023your}. 
Another study found that disclosing how a users’ data will be processed, another form of transparency, increased users’ WTA by at least 35\%~\cite{palinski2022paying}. 
However, their methodology makes it difficult to know if this effect is from this transparency, or simply priming participants to consider privacy—other studies have found that priming users to think about privacy reduces their chance of sharing data~\cite{john2011strangers,tsai2011effect}.

\subsection{Data donation for user generated data}

Broadly, there are two primary ways that researchers have historically gained access to user generated data to study behavior or audit platforms: (1) using web scraping, APIs, or other tools to collect data from platforms directly; and (2) obtaining privileged access to data through a partnership or commercial agreement with a company~\cite{breuerPracticalEthicalChallenges2020}. 
However, these approaches also carry significant drawbacks. For (1), public API access is being increasingly limited by online platforms~\cite{brunsAPIcalypseSocialMedia2019}. 
Also, data on "public" behavior scraped from the web may violate individuals’ privacy expectations and risks breaking websites’ terms of service~\cite{mancosuWhatYouCan2020}. 
For (2), financial or other collaborations with companies raise questions about privileged access and researcher independence~\cite{boyd2012critical}. Furthermore, none of these approaches address obtaining the individual consent of subjects whose data is being collected and used for research, raising serious ethical concerns~\cite{flickInformedConsentFacebook2016}.

One strategy that avoids some of these issues is to use data donation~\cite{lukitoEnablingIndependentResearch2023}. 
Data donation refers to individuals consensually sharing data for research or other purposes~\cite{prainsackDataDonationHow2019}. 
There are several ways to practically implement a data donation strategy. The emergence of data access rights in the EU and parts of the US has enabled a “download-upload” approach, where users download their data from a service and then submit that data to researchers~\cite{pfiffner2023leveraging, boeschotenFrameworkPrivacyPreserving2022, raziInstagramDataDonation2022}. 
Rather than using existing data download tools provided by platforms, users can also be empowered to collect their own data using custom software. For example, researchers have developed browser extensions that allow users to collect data on their browsing habits and then share that data with researchers, a form of data donation that avoids using platform-provided tools entirely~\cite{ricksDoesThisButton2022, wojcieszakAvenuesNewsDiverse2022}. 
In other cases where data exports are not available, data donation can also encompass less sophisticated forms of data, such as screenshots of applications that can then be turned into data usable by researchers~\cite{calacci2022bargaining}.

Although data donation is a promising strategy for researchers who wish to analyze user generated data, it includes its own set of challenges. Ethical and scientific standards for how to reliably run data donation studies have not yet been established~\cite{boeschotenFrameworkPrivacyPreserving2022, lukitoEnablingIndependentResearch2023, skatovaPsychologyPersonalData2019}, and there are few openly available technical solutions to collect and process donated data~\cite{araujoOSD2FOpenSourceData2022,boeschoten2022privacy}. 
Another significant barrier to using data donation in research is understanding why people might or might not donate their data in the first place. Studies that have focused on this question have found that a feeling of “social duty” and an understanding of the data are important factors that impact people’s decisions to donate~\cite{skatovaPsychologyPersonalData2019,pfiffner2023leveraging}. However, these studies investigating participant motivations have relied on online surveys that present users with hypothetical, rather than real, sharing scenarios, restricting their validity. Nevertheless, the limited work examining these motivations in real-world donation settings has validated that data transparency and focusing on consent are important design considerations for researchers~\cite{breuerUsercentricApproachesCollecting2023}.
We contribute to this burgeoning field by developing a data donation tool, that prioritizes consent, and using this tool to study the impact of data transparency and other incentives on motivating users to share data, in both real and hypothetical contexts.

\section{Survey and experiment design}
\label{sec:survey_design}

\begin{figure}
\centering
\includegraphics[width=0.8\textwidth]{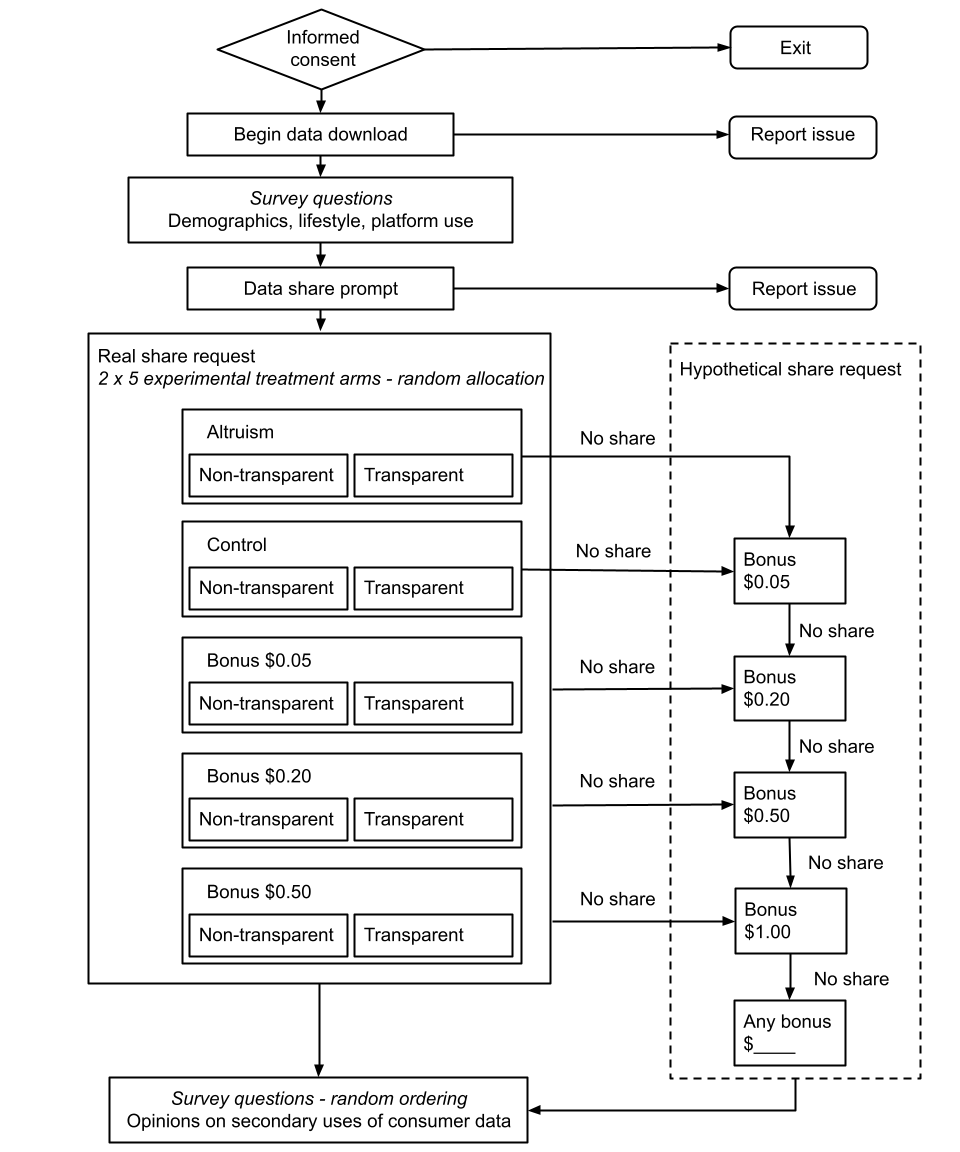}
\caption{Flowchart representing the survey. Each box represents a discrete section of the survey. Arrows represent movement from one section to another. “No share” indicates the flow if a participant declined to share within any treatment arm. Boxes within “Real share request” such as “Control” or "Bonus \$0.05" correspond to experimental treatment arms that participants were randomly assigned to. “Transparent” and “Non-transparent” boxes represent random assignment into either the transparent condition, where participants were shown their data before choosing to share, or non-transparent, where participants were shown only column names.}
\label{fig:survey_flowchart}
\end{figure}

This section describes the experimental study design and survey tool. 
The survey is represented at a high level in Figure~\ref{fig:survey_flowchart}. 
See the Appendix (\ref{SI:survey_tool}) for the full survey tool, including all text, questions, and answer options shown to participants.

The survey tool was designed to achieve multiple objectives. It was designed to collect Amazon users' purchase histories data while prioritizing informed consent. The tool also embedded an experiment designed to both test the impact of varying incentives and transparency levels on share rates, as well as to measure the "privacy paradox".
In addition, the survey collected information on participants' demographics, lifestyle and platform use, as well as participants' opinions on how purchases data, like the data collected by our tool, should be used.

\subsection{Ethics and informed consent}

This experiment was approved by the
MIT IRB (protocol \#2205000649).

The survey was designed to prioritize participant consent by allowing participants to opt in to sharing their Amazon purchases data. Care was taken to design a survey tool such that no purchases data left a participant's machine without their active consent. 
Participants were paid whether or not they chose to share their data, and this information was made clear to participants.

Before starting the study, participants were shown general information required by our IRB and provided informed consent in order to continue. 
They were then shown a clear outline of what the study would ask of them, namely that the study would guide them through downloading their purchases data, and later ask them to share it, and that PII and payment information would be stripped out. This page reiterated that they would be paid whether or not they shared the purchases data. It also told them that Amazon data they shared may be made public, warning "there is a chance that someone who knows a few of your past purchases may infer your identity". Participants had the option to exit without payment or continue to the study with informed consent (shown as "Informed consent" in Figure~\ref{fig:survey_flowchart}).

In addition to the base pay for participation, our study offered some participants (depending on their experimental treatment arm) monetary bonuses as incentives for data sharing.
Caution was taken around the potentially coercive tactic of offering monetary rewards to crowdworkers in exchange for their potentially sensitive data. After consultation with our IRB, we limited bonuses to \$0.50 to reduce the risk of bonuses having an adverse impact on more vulnerable participant populations.
In analysis, we use our study design to test whether the monetary bonuses have a larger impact on lower income participants. Regardless, given that we use small bonus amounts, researchers should take caution applying results from this test to inform the use of larger bonuses.

\subsection{Experiment design}
\label{sec:experiment_design}

\begin{figure}[htbp]
    \begin{adjustwidth}{-0.7cm}{}
\centering
\includegraphics[width=0.52\textwidth]{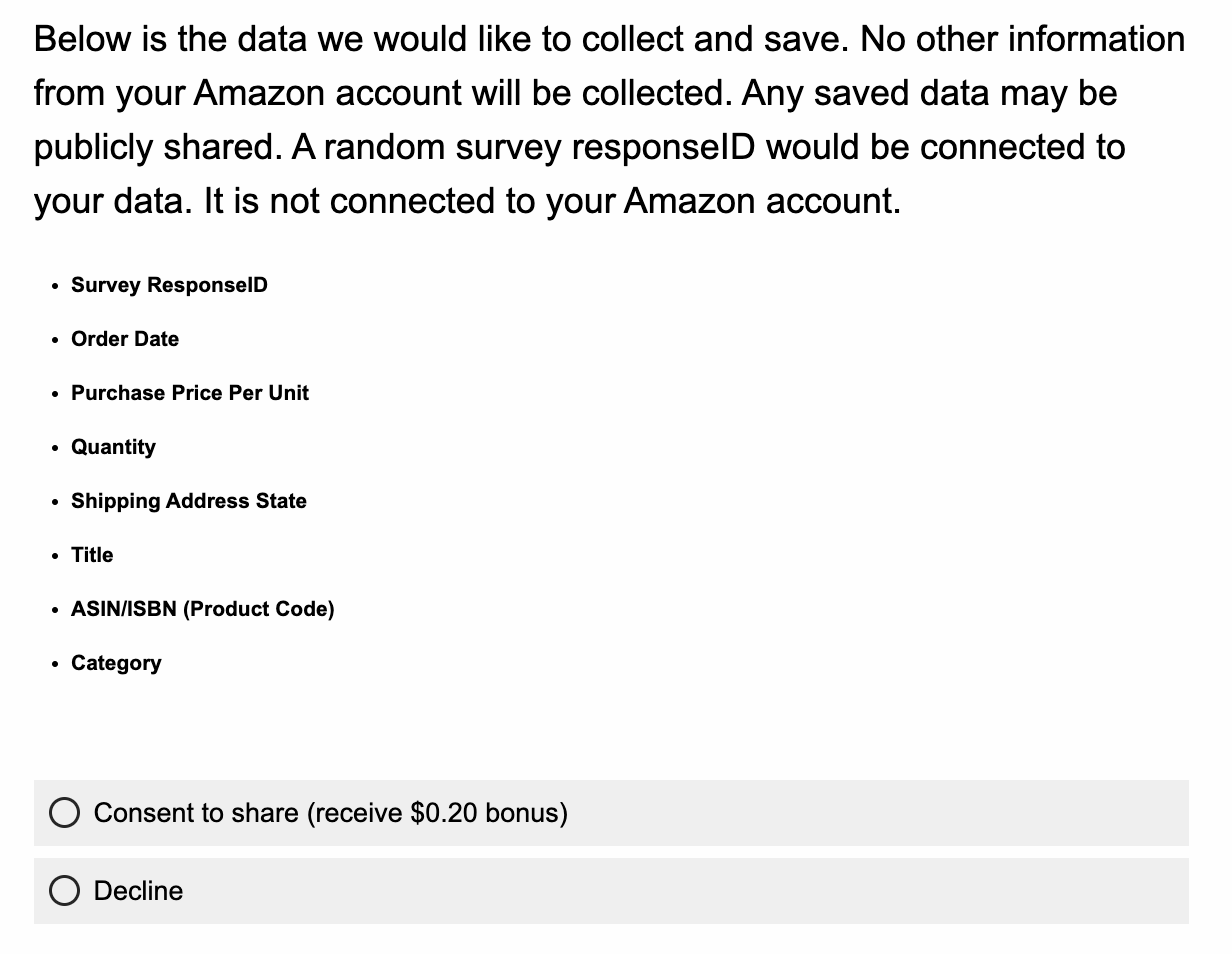}
\includegraphics[width=0.52\textwidth]{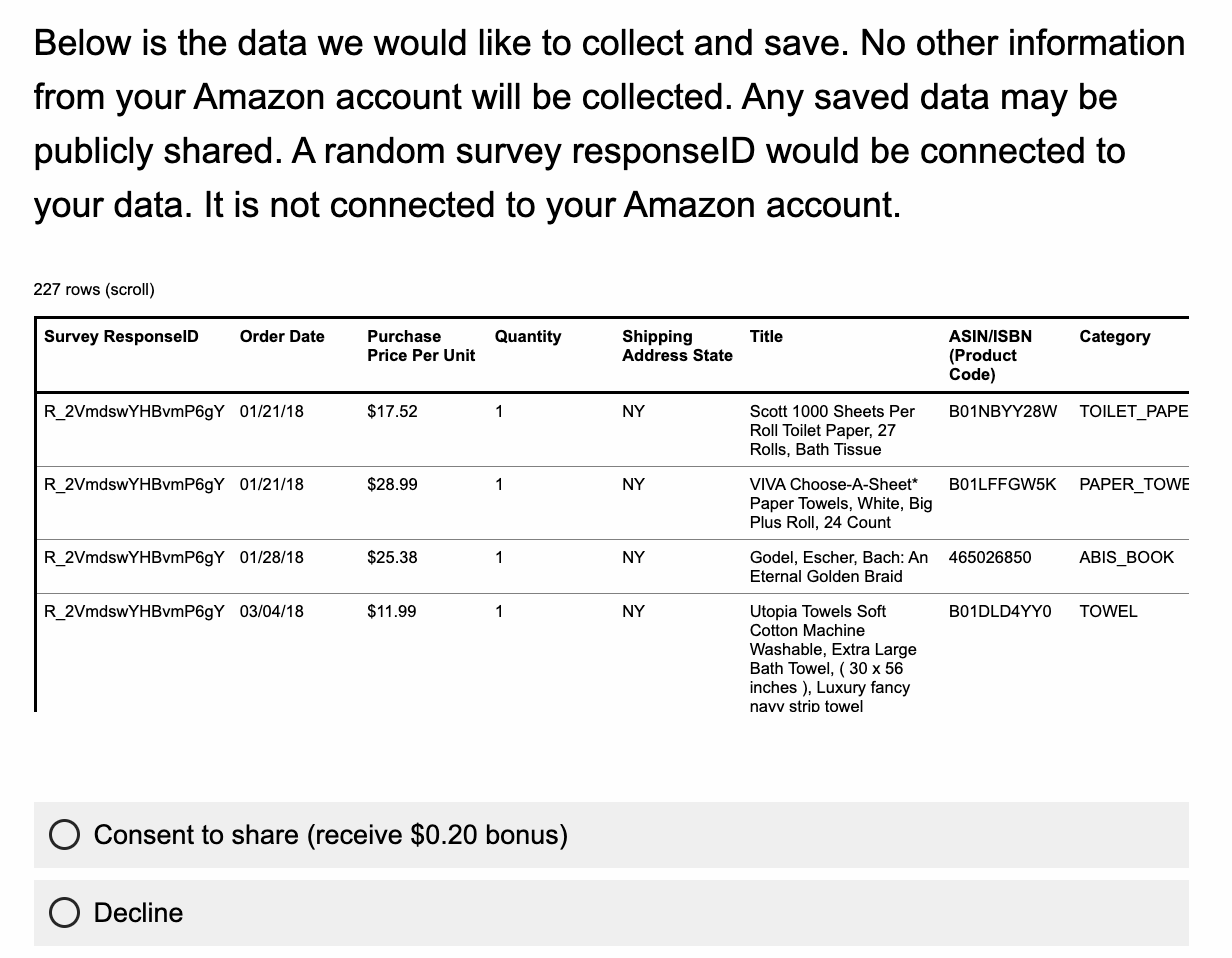}
\end{adjustwidth}
\caption{Screenshots from the survey's share request section. The experiment had a 2x5 factorial design, with 2 "transparency" treatments and 5 "incentives", with 10 total experiment arms. Shown is the experiment arm with the "transparent" and "\$0.20 bonus" treatments. Left: Interface before inserting Amazon data file. Right: Interface after inserting Amazon data file. Software within the browser stripped the data file to only include the data columns the survey text described to participants.
No data left participant machines unless they clicked "Consent to share". 
The interface for the "transparent" treatment presented participants with all rows and columns of data that would be collected, within a scrollable interface, before they chose to consent or decline to share. The "non-transparent" treatment only showed the data columns.}
\label{fig:share_prompt_example_20_showdata_short}
\end{figure}

Our experiment design allows us to measure three main constructs: (1) the effect of framing and monetary incentives on the likelihood that a participant chooses to share their data, (2) the effect of data transparency on share likelihood, and (3) the difference between real-world share rates and hypothetical share rates.

The experiment had a 2 $\times$ 5 factorial design, resulting in ten separate experimental treatment arms.
There were 5 different "incentive" treatments, and 2 different "transparency" conditions.
Participants were randomly assigned to an experimental arm with equal probability upon entering the survey.
The survey was identical for all participants except on one page of the survey where they made a choice whether or not to share their data ("Real share request" in Figure~\ref{fig:survey_flowchart}).
Upon reaching this page, participants had already downloaded (but not shared) their Amazon purchase histories data.
When prompted to share or decline to share their data, participants were presented with slightly different interfaces based on their experiment arm.

The 5 different "incentive" treatments included a control and three different bonus amounts of \$0.05, \$0.20, \$0.50. 
Participants in the bonus incentive arms were offered these bonuses in addition to the base pay for the survey, if they chose to share their data.
The fifth incentive treatment, referred to as the "altruism" incentive for brevity, added additional text that framed data sharing as an altruistic act: “...We are crowdsourcing data to democratize access to it as a public good and we are asking for your help..." See the Appendix (\ref{SI:survey}) for the full text for each treatment.
All experiment arms contained the text present in the control condition, which
explained exactly what data would be collected if they chose to share and that no other information from their Amazon account would be collected.
Additional text was added for the bonus and altruism treatments.

The two different transparency treatments included a "non-transparent" and "transparent" condition. In both treatments, participants were shown the names of the data fields that would be collected if they chose to share.
Participants in the "transparent" condition were also presented with all rows and columns of data that would be collected, within a scrollable interface, before they chose to consent or decline to share.

Figure~\ref{fig:share_prompt_example_20_showdata_short} shows an example of the interface participants saw.
It shows screenshots from the survey's share request section for the experiment arm with the "transparent" and "\$0.20 bonus" treatments, excluding the incentive specific text. (See Figure~\ref{fig:share_prompt_example_20_showdata} in the Appendix for an expanded view.)
Shown left is the interface before the participant inserted their Amazon data, while the right side shows after, where the data are presented for the participant to examine before choosing whether to share.
Software within the browser stripped the data file to only include the data columns the survey text described to participants and no data left participants' machines unless they clicked "Consent to share". 


Participants who declined to share were directed to a series of questions designed to measure the hypothetical amount they would have shared their data for, if any. 
These questions asked "Would you hypothetically consent to share your data for a bonus payment of \$X? (Since you already declined to share your data, your Amazon data will not be collected if you say yes)”.
The hypothetical bonus amounts (\$X) included the real bonus amounts (\$0.05, \$0.20, \$0.50) as well as \$1.00.

The amount that participants were first offered in this hypothetical scenario was determined by their original incentive treatment. For example, participants in the \$0.20 bonus condition who declined were asked if they would instead hypothetically share for a \$0.50 bonus. 
If a participant answered “Yes” they exited the hypothetical share request section of the survey. If they answered “No” they were offered the next highest hypothetical bonus amount. If a participant declined the hypothetical after being offered \$1.00, they were then asked "How much would you share your data for?" where they could write in a value or indicate ``I would not consent to share my data for any amount."
The arrows in Figure~\ref{fig:survey_flowchart} labeled “No share” indicate this user flow.

\subsection{Participant eligibility and prescreen}
Participants were eligible for the study if they met the following requirements: were U.S. residents, 18 years or older, had an active Amazon account they had used to make purchases since at least 2018, and could sign into this Amazon account during the survey.
These requirements were made clearly visible to prospective participants. We also used a prescreen survey to ensure eligibility and interest, where prescreen participants were informed of details of the main survey they might then participate in. The brief prescreen asked questions pertaining to each requirement and also included an attention check. It also tested if participants could access an “order history reports” page that would allow them to export a dataset of their Amazon order history, which would be used in the main study. Participants could also indicate they were not interested in accessing the page or participating in the main study.
Participants who passed the prescreen were then invited to participate in the main survey.

\subsection{Survey}

The survey involved participants using their Amazon accounts to download their purchase histories. This was done via an “order history reports” page provided by Amazon, which has since been taken offline\footnote{The page was accessible at \url{https://www.amazon.com/gp/b2b/reports}.}. Later in the survey ("Real share request" in Figure~\ref{fig:survey_flowchart}) participants were given the opportunity to share or decline to share a scrubbed version of this data.

After providing informed consent, participants were directed through the data download process. We asked participants to generate an export of their purchase history data starting from January 1st, 2018 to the current date they were completing the study (data were collected over the period of November 2022 to March 2023).
Amazon’s tool took a variable amount of time to process an order history download request. This unpredictable delay motivated us to design the survey in a way that enabled participants to answer other questions while waiting for their download to process. 

After beginning the download process, the survey asked participants questions about demographics, lifestyle, and platform use.
The demographic questions collected information about participants’ age, their self-identified race and ethnicity, their educational background, household income, gender identity, sexual orientation, and state of residence. We also asked participants for details about their Amazon usage, including how many others they shared their account with, how many people they considered to be in their “household”, and how often they typically ordered deliveries from Amazon. Other questions not examined in this paper included questions on behavioral health, such as alcohol and cigarette use and major life changes. 

Participants then entered the "data share prompt" section of the survey (see Figure~\ref{fig:survey_flowchart}) and were asked to finish downloading their report (which should then have finished processing). 
They were told they would next be asked to share their data and were reminded they would be paid whether they consented or declined to share. 
At both the initial data download and data prompt stage, participants had the opportunity to report an issue, which led to a separate survey flow where they described their issue with a screenshot.

Order history reports from the Amazon export are saved as CSV files, with a row for each item purchased. Our survey tool collected a specific subset of the CSV data columns, which contained no PII. These were: Order Date, Purchase Price Per Unit, Quantity, Shipping Address State, Title, ASIN/ISBN (Product Code), Category.
The CSV data contained other information that our tool did not collect, which included PII such as payment information and granular delivery address.

The page that prompted participants to share their data explicitly listed the data columns that our survey tool was requesting to collect, and stated that no other information would be collected (Figure~\ref{fig:share_prompt_example_20_showdata_short}).
This page prompted participants to insert the CSV file exported from Amazon into the browser web page. Software within the browser parsed the file, stripped out all data fields except the columns participants were told may be collected, and verified the file was valid.
At this point no data from the order history report left the participant's machine.
The reason the survey had all participants go through the data download and share prompt steps, whether or not they would share, was to make sure that participants declining to share represented a true choice not to share versus that they had not accessed their data.

Participants then saw language and an interface specific to the experiment arm they were assigned to, as described in Section~\ref{sec:experiment_design}. They then continued by clicking to either consent or decline to share. If they consented to share, the stripped data file was uploaded. If they declined to share, no data from the Amazon file was shared and participants continued to the hypothetical share request flow described in Section~\ref{sec:experiment_design}.

All participants were then asked to answer a series of questions about their opinions on how purchase history data, like the data they just downloaded, should be used by different parties. Participants were first shown a table displaying the same fields as in the data share request. Questions asked participants for a ``Yes"/``No"/``I don't" know response to a variety of questions such as ``Big companies currently collect and sell consumer purchase data. Do you think that small businesses should be able to access this data for free in order to help them compete with big companies?”. The list of questions from this section are shown in Figure~\ref{fig:data_use_results_summary}. This section also contained an attention check, randomly placed among these questions.
Finally, participants were provided a space to optionally insert free-text comments and were thanked for their time. 

\subsection{Software}
\label{sec:software}

The prescreen and main study surveys used Qualtrics survey software.
Without customization, Qualtrics software did not support our experimental design and privacy goals.
We developed custom software to integrate with the main study's Qualtrics survey. Our software integrations enabled showing the different interfaces in the share prompt, as determined by the experimental arms. The software was also developed to parse and handle the Amazon data file within the participants' machines and to ensure that no data were collected beyond the data columns shown to participants, and that the data did not leave participants' machines without their explicit consent.
The client-side software also validated the file provided by participants included the specified columns and contained rows of data representing at least two years.
All of the custom software is provided in an open source repository (see Data and Code Availability).

\section{Data and preprocessing}
\label{sec:data_and_preprocessing}

We consider the data collected by the survey described above as two distinct datasets: (1) survey data and (2) Amazon purchase histories data. The two datasets are connected by a Response ID randomly assigned to participants upon entering the survey.

We use "survey data" to mean the responses participants provided to survey questions, as well as the information from the experimental design, including whether or not they chose to share their data. The Amazon purchase histories data, which a portion of the participants chose to share, is maintained as a separate dataset. 
This paper analyzes the survey data and the following sections about preprocessing and the sample are about the survey data.
Readers can refer to the Appendix~(\ref{SI:amazon_data}) to learn more about the Amazon data collected.

All preprocessing and data analysis code are available via an open source repository (see Data and Code Availability below).
The repository also makes available the survey data analyzed in this paper, where raw data have been stripped of personally identifiable information (PII) and free text comments, corresponding to IRB guidelines.

\subsection{Data collection}
\label{sec:data_collection}

We collected data in a series of batches between November 2022 and March 2023.
We stopped collecting data on March 20, 2023 when Amazon took the Order History Reports page offline, which the survey tool depended on.

We restricted participant eligibility to U.S. residents 18 years or older who had an active Amazon.com account since at least 2018, and which they could log into during the survey. 
Recruitment was a two step process. Potential participants were first invited to a prescreen survey. Information about the main survey was made clear to potential prescreen participants. Prescreen participants with responses indicating they were both eligible and interested in the main survey were then invited to the main survey.
We presented  requirements to participants before they clicked through to participate, in both the prescreen and main survey. We offered prescreen participants \$0.35 for an estimated 1 minute survey (\$21/hr).
We offered participants \$1.50 for the main survey with an estimated 4-7 minute completion time. Some participants received additional bonus payments of \$0.05, \$0.20, \$0.50, as described in the survey design. 

We initially recruited participants via Prolific, an online survey recruitment platform (\url{https://www.prolific.co}). 
Due to a limited number of eligible U.S. participants on the Prolific platform, we also recruited using the CloudResearch platform in March 2023.
We collected 21,892 total prescreen responses, with 4,430 from CloudResearch. 6.2\% of prescreen participants failed the attention check (and were therefore deemed ineligible). 14,010 of the 21,892 total responses (64\%) indicated participants were both eligible and interested in the main survey. However, only a subset continued to participate in the main survey (see section~\ref{sec:data}).

\subsection{Preprocessing}
\label{sec:Preprocessing}

We processed the main survey data to provide a clean and publicly available dataset which is then analyzed in the following work.
To do this, we first excluded incomplete responses and responses with a failed attention check (less than 1\% failed the attention check in the main survey).
Since we recruited participants from multiple platforms, it was possible participants who work on both platforms could participate more than once. 
We did further preprocessing to remove duplicate responses, where we used the Amazon purchases data to identify duplicate purchase histories and drop corresponding responses from the survey data (200 responses were dropped).
We stripped the survey data of PII, including the participant IDs assigned by the survey recruitment platform which we used to pay the participants. Response IDs that were randomly assigned to participants upon entering the survey, and which cannot be used to link them to another platform, were retained. We also removed free text comments from the sample data to comply with IRB guidelines.

\subsection{Data} 
\label{sec:data}

\begin{table}[h]
\centering
\small
\begin{tabular}{lccc}
\hline
& \multicolumn{2}{c}{\textbf{Survey}} & \textbf{Census} \\
\hline
\textbf{Attribute} & \textbf{N} & \textbf{\%*} & \textbf{\%} \\
\hline
\multicolumn{4}{l}{\textbf{Gender}} \\
Female & 3132 & 50.9\% & 51.0\% \\
Male & 3020 & 49.1\% & 49.0\% \\
Other / Prefer not to say & 173 & - & - \\
\multicolumn{4}{l}{\textbf{Age}} \\
18 - 34 years & 3302 & 52.2\% & 29.4\% \\
35 - 54 years & 2369 & 37.5\% & 32.3\% \\
55 and older & 654 & 10.3\% & 38.3\% \\
\multicolumn{4}{l}{\textbf{Household income}} \\
Less than \$50,000 & 2298 & 37.1\% & 35.5\% \\
\$50,000 - \$99,999 & 2315 & 37.4\% & 30.3\% \\
\$100,000 or more & 1582 & 25.5\% & 34.1\% \\
Prefer not to say & 130 & - & - \\
\multicolumn{4}{l}{\textbf{Education level}} \\
High school or GED or less & 2340 & 37.4\% & 65.2\% \\
Bachelor's degree & 2834 & 45.3\% & 22.1\% \\
Graduate or professional degree & 1086 & 17.3\% & 12.7\% \\
Prefer not to say & 65 & - & - \\
\multicolumn{4}{l}{\textbf{Race}} \\
White & 4825 & 76.3\% & 75.8\% \\
Asian & 551 & 8.7\% & 6.1\% \\
Black or African American & 440 & 7.0\% & 13.6\% \\
Other or mixed & 509 & 8.0\% & 4.5\% \\
\multicolumn{4}{l}{\textbf{Online purchase frequency}} \\
Less than 5 times per month & 4081 & 64.5\% & - \\
5 - 10 times per month & 1760 & 27.8\% & - \\
More than 10 times per month & 484 & 7.7\% & - \\
\hline
\multicolumn{4}{l}{* Survey \% excludes counts for "Other" and "Prefer not to say"}\\
\hline
\end{tabular}
\caption{Summary of participant attributes.}
\label{tab:participant_attributes}
\end{table}

The survey sample includes 6,325 participants, after preprocessing. 
Table \ref{tab:participant_attributes} shows participant attributes as they are used in the analyses. To simplify analysis and ease interpretation, we group some attributes. For example, while we collected more detailed age data, we aggregate to three groups for analysis. Table \ref{tab:participant_attributes_detailed} in the Appendix shows participant attributes as they were collected. Given that eligible participants were 18 years or older, wherever possible we compare the sample data to census data for the 18 or older population.

For gender, when we restrict sample data to the Male/Female binary by excluding responses for ``Prefer not to say" and ``Other" there is an exact match of 51\% vs 49\% to U.S. census data for the 18 and older population~\cite{census2022DP05}. 
Our sample demonstrates an important age bias, under-representing older participants and over-representing younger participants\cite{census2022agesex}, which is a limitation. 
The sample slightly under-represents higher-income households, while slightly over-representing lower and middle-income households \cite{census2022asec}. 
Similarly, our sample over-represents individuals with a bachelor's degree or greater level of education and under-represents those with a high school education or less \cite{census2022asec}. 
Our survey allowed the selection of multiple racial identities. For analysis, we grouped participants who identified as a single race (White, Asian, or Black) separately, and grouped those who answered "American Indian/Native American or Alaska Native", ``Native Hawaiian or Other Pacific Islander", "Other", or who reported 2 or more races, as "Other or mixed". The sample's racial distribution is overall similar to census data but slightly under-represents Black participants \cite{census2021sr11h}.  
Our sample's geographic distribution is highly correlated with the U.S. population by state (Pearson correlation of 0.981, p-value<0.001)~\cite{census2022SCPRC-EST2022-18+POP}, with exceptions like the absence of participants from Puerto Rico and an imbalance in representation from California, Texas, and Pennsylvania. Participation by state can be found in Appendix Table~\ref{tab:participant_state}.

In addition to sample demographics, we consider potential bias due to (1) non-response bias and (2) attrition within the main survey.
For (1) we use the last question of the prescreen survey and leverage participant demographics that were included in the prescreen survey data from Prolific.
The last question of the prescreen gauged both eligibility and interest in the main study, where participants could indicate they were not interested (opt out). 
We analyze a potential relationship between demographics and likelihood to opt out.  Details are in the Appendix (\ref{SI:nonresponse_bias}).
Regression results show older participants were more likely to opt out (p<0.05).
This likely exacerbated the under-representation of older participants in the sample, as shown above. 
For (2) we assessed whether the experiment arm impacted participants’ likelihood to complete the survey. We confirm that this was not the case using a chi-squared test of homogeneity which shows that there was no significant variation in completion rates across treatment arms (see Appendix \ref{SI:attrition_bias}).

\section{Experiment results and data sharing behavior}
\label{sec:experiment_results}

\begin{figure}
\centering
\includegraphics[width=0.5\textwidth]{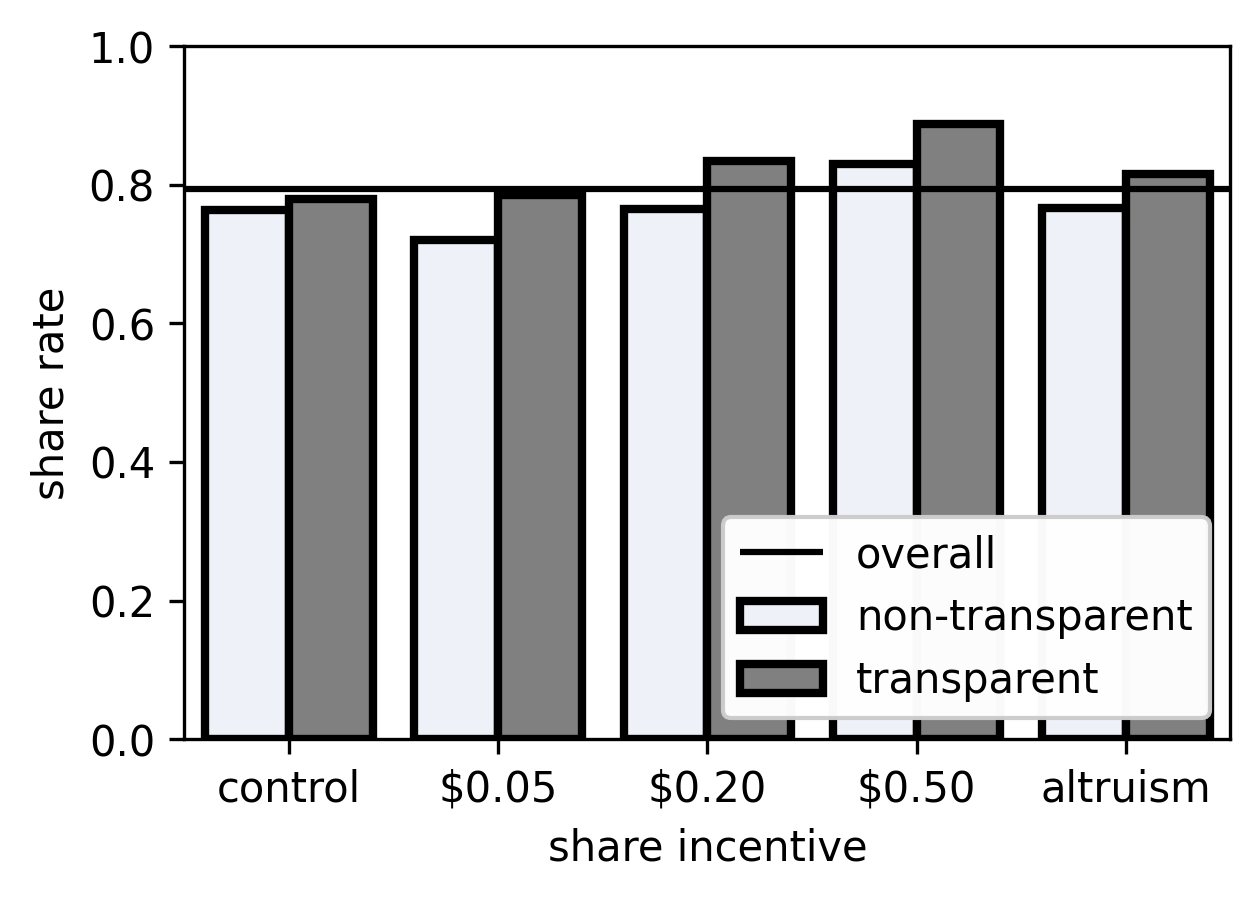}
\caption{Share rates by experiment arm. Participants were randomly assigned to an experiment arm in a 5x2 experimental design, with 5 "incentives", shown on the x-axis. In the transparent condition, when prompted to share their data, participants were shown a table with all data that would be shared, while in the non-transparent condition, participants were only shown the data column headers.}
\label{fig:overall_share_rates}
\end{figure}

A total of 6,325 participants completed the survey.
Overall, across all experiment arms, 79.4\% of participants shared their data.
We refer to the portion of participants who chose to share their data as the share rate.
The share rates varied by experiment arm which is shown at a high level in Figure~\ref{fig:overall_share_rates}.
The number of participants and share rates for each experiment arm are in Appendix Table~\ref{tab:overall_share_rates}.

Here we make high level observations which are then quantified and tested for significance, while controlling for participant level attributes.
First, Figure~\ref{fig:overall_share_rates} illustrates a clear positive monotonic relationship between the size of a bonus incentive and share rates. Further analyses identify a linear relationship between the additional bonus amount and share rates.
We note that the control group was unaware of monetary rewards and participants offered a small bonus of \$0.05 did not exhibit higher share rates than the control group.

Second, for every incentive treatment, participants who were shown their data (i.e. in the transparent data condition) were more likely to share. 
In particular, participants in the altruism experiment arm (where participants were told why their data would be helpful) and who were also shown their data (transparent condition) shared more often than participants in the control. 
Furthermore, participants in the experiment arm with a \$0.20 bonus and who were shown their data (transparent condition) shared more often than those in the \$0.50 bonus arm who were not shown their data.
These observations suggest that transparency can be valuable to researchers conducting crowd sourcing campaigns, when compared to extra marginal payment. Namely, providing clarity in what data would be collected from participants may be both more ethical and cost efficient for researchers.

\subsection{Regression analysis methods}

We use logistic regression models to quantify the observed effects, test their significance, and evaluate the impact of participant demographics and other covariates on the likelihood of data sharing. The dependent variable in each model is a boolean ``share'', indicating whether a participant chose to share their data. The ``transparent'' variable is a boolean indicating whether participants were in the transparent condition (shown their data). Finally, $\vec{\text{attributes}}$ represents a vector of participant characteristics modeled as categorical variables. These include gender, age, household income, education, race, and purchase frequency, as shown in Table~\ref{tab:participant_attributes}, where responses with "Prefer not to say" or gender "Other" are excluded.

We use two separate models to measure the impact of the experimental treatments. 
Model 1 is represented by Equation~\ref{eq:model_1} and model 2 is represented by Equation~\ref{eq:model_2}. 
Their different definitions allow for different types of comparisons.
\begin{equation}
\text{{share}} = \text{transparent} + \text{{incentive}} + \vec{\text{attributes}}
\label{eq:model_1}
\end{equation}
\begin{equation}
\text{{share}} = \text{control} + \text{{altruism}} + \text{bonus}_{\$0.05} + \ldots + (\text{transparent } \& \text{{ bonus}}_{\$0.50}) + \vec{\text{attributes}}
\label{eq:model_2}
\end{equation}

Model 1 isolates the effects of the incentives versus transparency treatments. 
Furthermore, it allows isolating potential interaction effects between treatments and covariates. 
In particular, we extend the model and run an additional regression in order to test for interaction effects between the incentive and household income variables by adding a term to Equation~\ref{eq:model_1}: incentive~$\times$~household\_income.
We do this to test a potential concern where the monetary incentives could have a greater effect on lower income participants.
In Equation~\ref{eq:model_1}, ``incentive'' is a categorical variable.

Model 2 allows directly comparing and quantifying effect sizes across the 10 experiment arms, relative to the control incentive without the transparent treatment.
Each of the combinations of the transparency and incentive treatments are modeled as dummy variables as shown in Equation~\ref{eq:model_2}, where  the ``transparent'' boolean is only included for the groups with the transparent data treatment.

\subsection{Results}

\begin{table}[h]
\centering
\small
\begin{tabular}{lccc}
\hline
\textbf{Predictor} & \textbf{B (log odds)} & \textbf{Odds ratio} & \textbf{95\% CI for Odds Ratio} \\
\hline
Intercept & 0.999*** & 2.715 & [2.140, 3.446] \\
Transparent\textsuperscript{1} & 0.374*** & 1.453 & [1.277, 1.654] \\
\multicolumn{4}{l}{Incentive (Reference: control)} \\
altruism & 0.039 & 1.040 & [0.854, 1.266] \\ \$0.05 & -0.152 & 0.859 & [0.708, 1.043] \\ \$0.20 & 0.08 & 1.083 & [0.886, 1.323] \\ \$0.50 & 0.584*** & 1.793 & [1.439, 2.234] \\
\hline
\multicolumn{4}{l}{\textbf{Participant attributes}} \\
\hline
\multicolumn{4}{l}{\textbf{Gender (Reference: Male)}} \\
Female & 0.367*** & 1.443 & [1.266, 1.644] \\
\multicolumn{4}{l}{Age (Reference: 35 - 54 years)} \\
18 - 34 years & -0.176* & 0.839 & [0.728, 0.966] \\
35 - 54 years & -- & -- & -- \\
55 and older & -0.062 & 0.940 & [0.745, 1.184] \\
\multicolumn{4}{l}{\textbf{Household income (Reference: \$50,000-\$99,999)}} \\
Less than \$50,000 & 0.145 & 1.156 & [0.989, 1.351] \\
\$50,000 - \$99,999 & -- & -- & -- \\
\$100,000 or more & 0.046 & 1.047 & [0.888, 1.235] \\
\multicolumn{4}{l}{\textbf{Education (Reference: Bachelor's degree)}} \\
High school or GED or less & 0.16* & 1.173 & [1.011, 1.362] \\
Bachelor's degree & -- & -- & -- \\
Graduate or professional degree & 0.127 & 1.135 & [0.945, 1.364] \\
\multicolumn{4}{l}{\textbf{Race (Reference: White)}} \\
Asian & -0.561*** & 0.571 & [0.462, 0.704] \\
Black & -0.072 & 0.931 & [0.719, 1.206] \\
Other or mixed & 0.021 & 1.021 & [0.796, 1.310] \\
\multicolumn{4}{l}{\textbf{Purchase frequency (Reference: 5-10 times/month)}} \\
Less than 5 times per month & -0.035 & 0.966 & [0.832, 1.122] \\
5 - 10 times per month & -- & -- & -- \\
More than 10 times per month & -0.178 & 0.837 & [0.649, 1.080] \\
\hline
\multicolumn{4}{l}{N = 5995} \\
\multicolumn{4}{l}{Pseudo R-squared = 0.028} \\
\hline
\end{tabular}
\caption{Model 1 results: Likelihood to share as a function of incentive, data transparency (whether data are shown), and participant level attributes.
\newline
\textsuperscript{1}The non-transparent data treatment is used as the reference variable. Other reference variables are indicated in the table alongside their corresponding category name.
}
\label{tab:model1_results}
\end{table}

\begin{table}[h]
\small
\centering
\begin{tabular}{lccc}
\hline
\textbf{Experiment arm} & \textbf{B (log odds)} & \textbf{Odds ratio} & \textbf{95\% CI for Odds Ratio} \\
\hline
Intercept: Control (Non-transparent) & 1.065*** & 2.902 & [2.226, 3.783] \\
altruism & -0.01 & 0.990 & [0.758, 1.294] \\ \$0.05 & -0.211 & 0.810 & [0.623, 1.053] \\ \$0.20 & -0.007 & 0.993 & [0.758, 1.300] \\ \$0.50 & 0.398** & 1.490 & [1.114, 1.991] \\
\hline
Transparent & & & \\
\hline
Control & 0.223 & 1.250 & [0.945, 1.653] \\
altruism & 0.318* & 1.374 & [1.036, 1.823] \\ \$0.05 & 0.139 & 1.149 & [0.870, 1.518] \\ \$0.20 & 0.407** & 1.503 & [1.121, 2.014] \\ \$0.50 & 1.053*** & 2.865 & [2.053, 3.999] \\
\hline
Note: Participant attributes omitted for brevity & & & \\
\hline
\end{tabular}
\caption{Abbreviated set of regression results for model 2. Shows effects of the 10 experiment arms: the relative impact of the incentive treatments with and without the transparent data treatment. 
\newline
Note: *p<0.05; **p<0.01; ***p<0.001.}
\label{tab:model2_results_abbreviated}
\end{table}

Results from Model 1 are presented in Table~\ref{tab:model1_results}. Showing participants their data had a statistically significant and substantial positive effect on share rates - participants in the transparent condition were nearly 1.5 times as likely to share (OR=1.45, 95\%CI [1.27, 1.65], p<0.001). Providing a large enough monetary incentive (\$0.50) also significantly increased the share rate (OR=1.79, 95\%CI [1.43, 2.23], p<0.001).

These results also show how participant demographics play a role.
Female participants were more than 1.4 times as likely to share than their male counterparts (OR=1.44, 95\%CI [1.26, 1.64], p<0.001).
Participants with lower education levels also showed a higher likelihood to share. The effect size found here is small, however our sample under-represents this population and more work should be done to explore whether education and data literacy impact share rates, given a lack of data literacy could result in a disproportionate and negative impact on lower education populations in other contexts. 
In addition, the results show younger participants and those identifying as Asian were less likely to share.

To investigate the potential differential impact of incentives across household income levels, we extended the model to include an interaction term ``incentive $\times$ household income''.  Results are provided in Appendix Table~\ref{tab:model1_income_incentive_interaction}.
We find no evidence that incentives had a different impact by household income group, as the effects of the interaction terms are not significant (p>0.05). In addition, the significance and direction of other variables in the model did not change.

Table \ref{tab:model2_results_abbreviated} shows results for Model 2, capturing only the effects of the experiment arms and omitting the participant-level covariates for brevity. Full regression results for Model 2 are in Appendix Table~\ref{tab:model2_results_full}. We note the same covariates are significant for Model 1 and Model 2 and they are consistent in sign.
These results allow directly comparing how the different combinations of experimental treatments impacted share rates relative to the control group when data were not shown (i.e. relative to incentive control and non-transparent condition).
81.6\% of participants in the group with the altruism incentive and transparent data treatment shared data versus 76.5\% in the control.
Monetary incentives alone also played an important role.
Participants in the group offered a \$0.50 bonus, yet not shown their data, shared 83.0\% of the time and this share rate was matched by the group offered the smaller bonus of \$0.20 who were in the transparent data condition (83.4\%).
The group offered both the largest bonus amount of \$0.50 with the transparent data condition shared 88.8\% of the time.
These results underscore the significant role of transparency in building trust and encouraging participation, as well as demonstrate how transparency can be a cost effective strategy to improve data sharing.

In the following analyses we also identify a linear relationship between bonus incentive amounts and share rates.

\section{Measuring the privacy paradox: Real versus hypothetical impacts of monetary bonuses}
\label{sec:measure_privacy_paradox}

In order to measure differences in real versus hypothetical data sharing contexts, our experiment asked participants who did not consent to share their data whether they would hypothetically share for a (larger) bonus incentive.
Participants were asked about incrementally larger amounts up to \$1 until they said yes.
If none of these amounts were sufficient, participants were asked to specify an amount for which they would hypothetically consent to share their data or to indicate that they would not do so for any amount (see Section~\ref{sec:survey_design} for details).
Of the N=6325 participants, 641 wrote in amounts. These write-in values are not used in the following analysis and are described in the Appendix (see \ref{appendix:write_in_amounts}).
330 participants (5.22\%) said they would not consent to share their data for any amount.

\begin{figure}
\centering
\begin{minipage}{0.5\textwidth}
  \centering
  \includegraphics[width=1\linewidth]{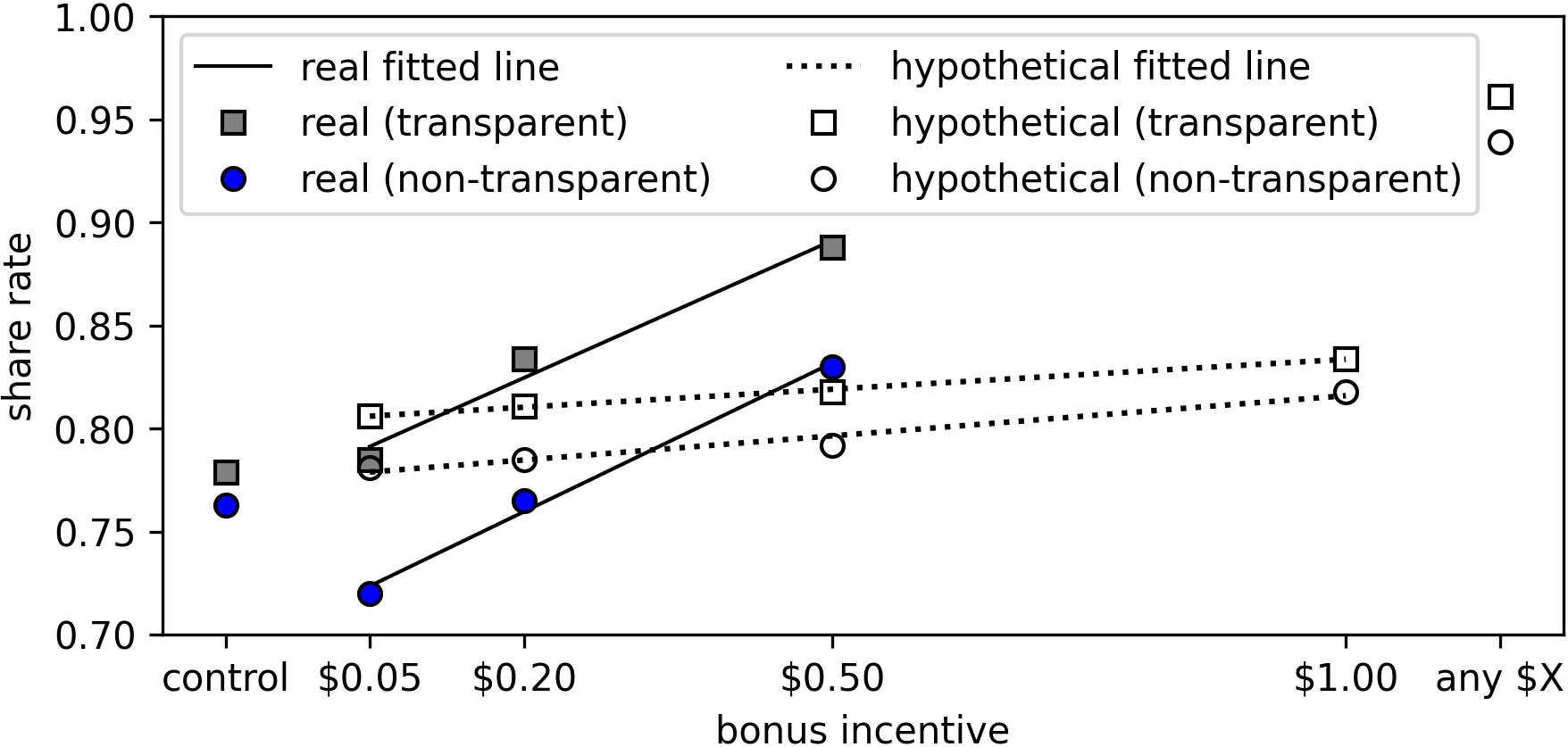}
\end{minipage}%
\begin{minipage}{0.5\textwidth}
  \centering
  \includegraphics[width=0.93\linewidth]{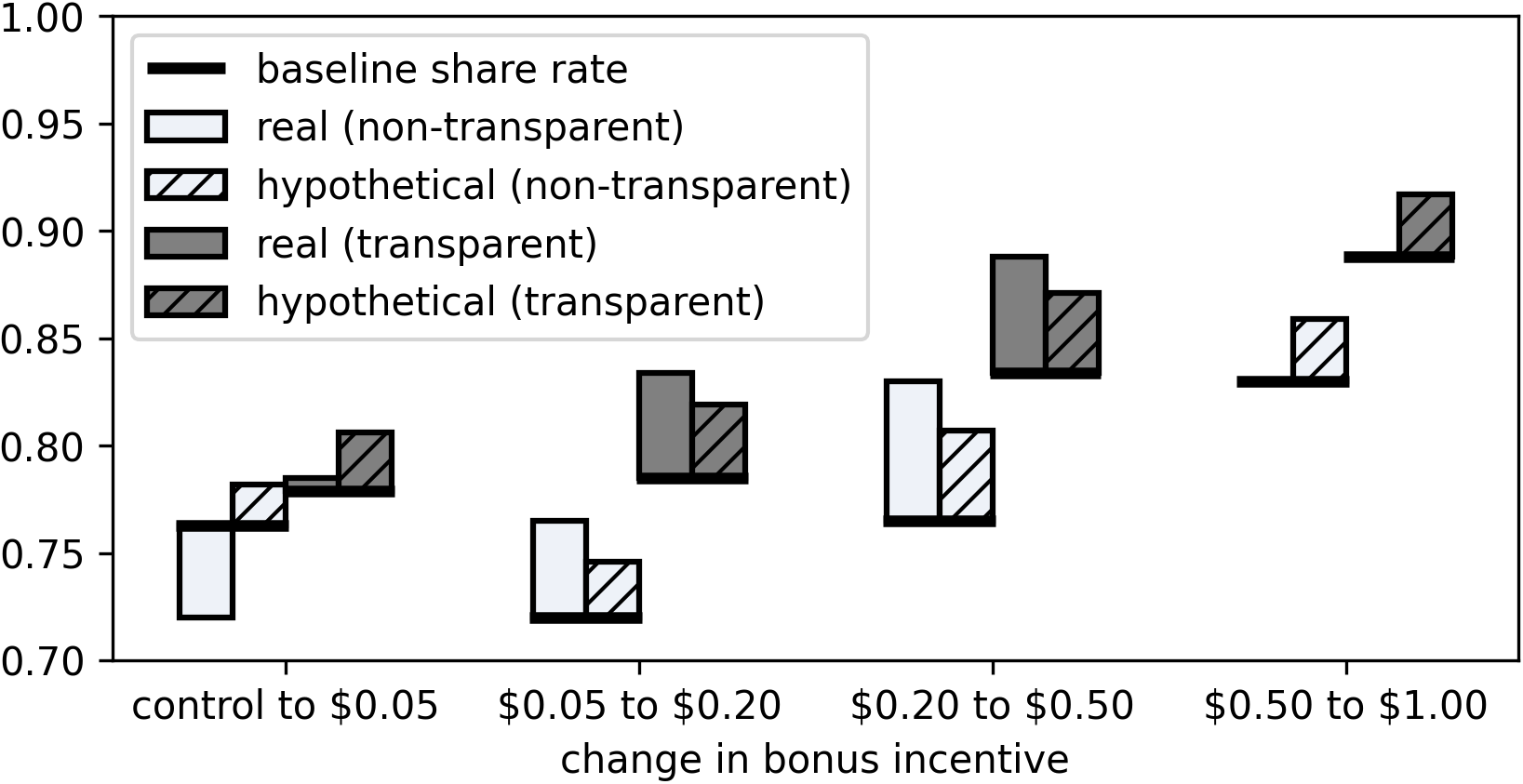}
\end{minipage}
\caption{Share rates for participants offered real versus hypothetical bonuses to incentivize sharing. Transparent versus non-transparent indicates whether participants were in a treatment group where they were shown their data when prompted to share (transparent). 
Left: Hypothetical share rates are computed from the control group as the cumulative portion of participants who agreed to share their data for less than or equal to a given bonus amount. Bonus amounts (\$0.05, \$0.20, \$0.50) are spaced on the x-axis corresponding to their values, illustrating a linear relationship between the dollar amount offered and share rates. "Real" share rates reflect data from experiment groups offered real monetary incentives. Note hypothetical shares were added to real shares in the control, resulting in a visibly higher intercept for hypothetical share rates -- only the slope should be interpreted.
Right: Change in share rate when comparing a given bonus incentive amount to the next smaller amount. The x-axis represents the change in incentive. For instance, "\$0.05 to \$0.20" labels the measured change in share rate for participants who were presented with a \$0.20 bonus incentive, using the \$0.05 bonus incentive as a baseline.}
\label{fig:real_vs_hypothetical}
\end{figure}

This section analyzes the trends, similarities, and differences in participants' stated willingness to share their data for a hypothetical monetary incentive versus their actual willingness to share when offered a real opportunity to earn a monetary bonus.
Two different strategies are used below. The first compares the impact of incremental additions in the real versus hypothetical bonus incentives. 
The second more directly compares the change in share rate going from one incentive to the immediately higher bonus amount, for the real versus hypothetical cases.
Together, these analyses help measure the extent to which the privacy paradox---the divergence between expressed and actual privacy behaviors---manifests in our study. 
More generally, this unique experiment data adds to the privacy paradox literature by quantifying differences in these privacy behaviors and exposing similarities, helping to evaluate the reliability of hypothetical data sharing contexts as a predictor for real-life actions.

\subsection{Comparing hypothetical versus real share rates based on the control}

This section evaluates impact of incremental additions in the real versus hypothetical bonus incentives.
Here, we compute the hypothetical impact by only using participants in the control group who initially declined to share their data (N=639 for the transparent data condition, N=636 for the non-transparent condition).
As a reminder: participants in the control group who declined to share were then asked if they would hypothetically share for a bonus amount of \$0.05, then \$0.20, \$0.50, \$1.00 or any amount, which they could write in. If they said yes, they were not shown a larger amount.
We compute the hypothetical share rate for each bonus amount as the portion of participants who said they would share for this amount or any lower value.
We compare this to the share rates for participants who were in the experiment arms such that they were given the real (not hypothetical) opportunity to share for these amounts.
This is shown in Figure~\ref{fig:real_vs_hypothetical}~(left), where the monetary rewards are spaced on the x-axis consistent with their values, and a line fit from a least squares regression is superimposed.
Note we do not consider the control condition as a \$0 incentive since participants who saw the control were unaware of the possibility of monetary rewards for data sharing.

A positive linear relationship between monetary reward and share rate can be observed. Moreover, this linear relationship is present for both the real and hypothetical sharing contexts and is consistent across the transparent and non-transparent data conditions.
Furthermore, the fact that share rates are higher in the transparent condition holds true in the hypothetical context.
An important difference is that the impact of additional monetary rewards is notably larger (a steeper slope) in the real versus hypothetical context.

We quantify these observations with a least squares linear regression for each set of data points: real (transparent condition), real (non-transparent condition), hypothetical (transparent condition), and hypothetical (non-transparent condition), using the equation:
\begin{equation}
\text{{share rate}} = \text{{intercept}} + \beta \times \text{{amount}}
\label{eq:real_vs_hypo_regression}
\end{equation}
where ``amount'' corresponds to the values \$0.05, \$0.20, \$0.50, and \$1.00 in the hypothetical scenario.

All data and estimated values used in the regression are available in Appendix Table~\ref{tab:real_vs_hypothetical_incremental}.

The results\footnote{Note the coefficients are scaled to \$1 for readability, while our bonus amounts were all less than \$1.} estimate the impact of an additional \$1 bonus on the share rate ($\beta$). For the real sharing contexts, the estimated $\beta$ values are 0.240 (Pearson r=0.996) and 0.222 (Pearson r=0.983) for the non-transparent and transparent conditions, respectively. 
In contrast, for the hypothetical sharing contexts, the estimated $\beta$ values are 0.039 (Pearson r=0.983) and 0.029 (Pearson r=0.998) for the non-transparent and transparent conditions, respectively. 
For each transparency condition, the estimated impact of additional bonus payment is more than 6 times for the real versus hypothetical scenario.

\subsection{Analyzing change in share rates due to real versus hypothetical incentives using all experimental arms}

We conduct a more direct comparison of the changes in share rates when progressing from one incentive to the next higher amount in both real and hypothetical contexts.

For a given bonus incentive $I_{i}$ the real share rate for the next smaller incentive, $I_{i-1}$, is used as a baseline. 
The real change in the share rate for incentive $I$ is computed by subtracting this baseline from the share rate for $I$. 
For example, the change in the real share rate when moving from a \$0.05 to a \$0.20 bonus is calculated by using the real \$0.05 share rate as a baseline and subtracting this value from the \$0.20 share rate. 

To compute the hypothetical changes in share rates for each incentive amount $I_{i}$, we restrict data to the $n_{i-1}$ participants in the experiment arm with the immediately smaller incentive $I_{i-1}$. 
Participants in that group who did not consent to share were then asked whether they would hypothetically share for the hypothetical incentive amount $I_{i}$. 
We add the number of participants who said yes to the number of real shares for the previous amount, considering these the participants who would hypothetically share for $I_{i}$ or less.
We divide this sum by the number of $n_{i-1}$ participants to compute the hypothetical share rate for incentive $I_{i}$.
The hypothetical change in share rate is then this hypothetical share rate minus the baseline real share rate for incentive $I_{i-1}$. 
Data tables for these computations are shown in the Appendix (see Table~\ref{tab:real_vs_hypothetical_direct}).
Results are shown in Figure~\ref{fig:real_vs_hypothetical}~(right). 


The results shown in Figure~\ref{fig:real_vs_hypothetical}~(right) align with the previous analysis that measured the incremental impact of additional bonus amounts, but where computation of hypothetical share rates was confined to the control experiment arm. 
Both Figure~\ref{fig:real_vs_hypothetical}~(right) and Table~\ref{tab:real_vs_hypothetical_direct} show that increased monetary incentives had a higher impact for the real versus hypothetical contexts.

\subsection{Summary of results}

The analysis reveals consistent trends across both real and hypothetical data-sharing contexts.
Across the real and hypothetical contexts, transparency positively impacts share rates. Furthermore, a linear correlation is observed between bonus amounts and share rates, for all of the real and hypothetical sharing contexts. This finding suggests that hypothetical scenarios could be effectively employed to gauge relationships between monetary incentives and data-sharing practices.

However, this analysis exposes important differences between the real and hypothetical contexts. Namely, the data-sharing rates computed for hypothetical scenarios consistently fall short of those seen in real scenarios. This suggests that when participants are presented with a hypothetical bonus incentive, they are less likely to declare a willingness to share their data compared to the actual behaviors observed when they are genuinely given the opportunity to earn a bonus. In particular, the incremental impact of additional monetary incentives on share rates was estimated to be 6 times higher for real versus hypothetical sharing contexts.

This phenomenon aligns with the existing literature on the ``privacy paradox", a concept that suggests individuals tend to claim they are less inclined to share their data than their actual behaviors might indicate. However, given the unique experiment design and large sample size, this study is able to quantify this difference more directly than previous work.

\section{Perceptions on data use}

\begin{figure}
\centering
\includegraphics[width=\textwidth]{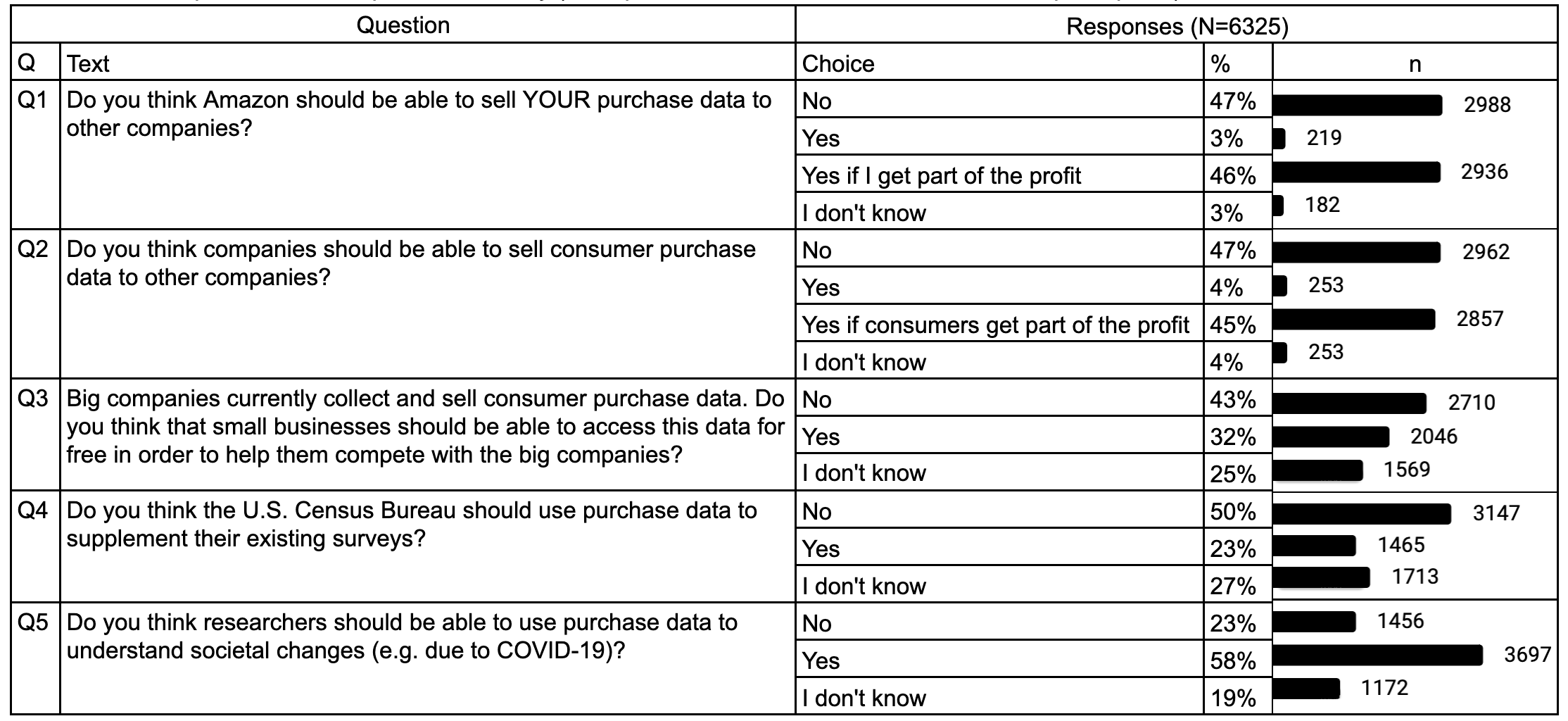}
\caption{Data use survey questions and summary of results. Question identifiers (Q1-Q5) are used for clarity and reference for the analyses. Questions were presented in random order to participants.}
\label{fig:data_use_results_summary}
\end{figure}

At the end of the survey participants were asked five questions regarding their views on how purchases data, like that collected by the survey, should be used. This is described in Section~\ref{sec:survey_design}.
The questions and resulting answers are shown in Figure~\ref{fig:data_use_results_summary}. 
Note the five questions are numbered as Q1, Q2, …, Q5, in order to add clarity to the following analyses and discussion of results. The questions were presented without numbers in random order to participants.

Q1 and Q2 asked participants their opinions on the use of purchase data by Amazon and other companies, respectively, and framed the question differently:
Q1 asked whether Amazon should be about to sell YOUR purchase data while Q2 asked more generally whether companies should be able to sell consumer data (Q2).
The results are nearly identical, where 47\% of participants answered "No".
While only 3-4\% answered Yes, more than 45\% answered "Yes if I/consumers get part of the profit". Only 3-4\% answered "I don't know".

Q3 sought the participants' opinions on whether smaller companies should have access to data collected by larger companies to maintain competitiveness. The responses to this question were mixed across "Yes/No/I don't know".

A stark contrast was observed between the responses to Q4 and Q5 which focused on data usage by the U.S. census and researchers.
50\% of participants expressed disapproval towards the U.S. Census Bureau utilizing purchase data to supplement their surveys.
Yet a majority (58\%) approved of researchers using such data.

To better understand the variation in responses, we use regression analyses to help reveal relationships between participant level attributes and responses to these questions.

\subsection{Regression analysis}

\begin{table}[]
\small
\begin{tabular}{lccccc}
\hline
\textbf{} & \textbf{Q1} & \textbf{Q2} & \textbf{Q3} & \textbf{Q4} & \textbf{Q5} \\ 
\textbf{Predictor} & \textbf{Odds Ratio} & \textbf{Odds Ratio} & \textbf{Odds Ratio} & \textbf{Odds Ratio} & \textbf{Odds Ratio} \\ 
\hline
Intercept & 0.466*** & 0.477*** & 0.264*** & 0.171*** & 0.73** \\
share & 3.028*** & 2.795*** & 3.451*** & 2.833*** & 3.829*** \\ 
\multicolumn{6}{l}{\textbf{Gender\textsuperscript{1}}} \\
Female & 0.524*** & 0.522*** & 1.142* & 0.572*** & 0.942 \\ 
\multicolumn{6}{l}{\textbf{Age\textsuperscript{2}}} \\
18 - 34 years & 1.529*** & 1.431*** & 1.295*** & 1.732*** & 1.338*** \\
55 and older & 0.578*** & 0.578*** & 0.874 & 0.717*** & 0.777* \\ 
\multicolumn{6}{l}{\textbf{Household income\textsuperscript{3}}} \\
Less than \$50,000 & 1.117 & 1.076 & 0.954 & 0.999 & 1.13 \\
\$100,000 or more & 1.014 & 1.094 & 0.855 & 1.233* & 1.183 \\ 
\multicolumn{6}{l}{\textbf{Education\textsuperscript{4}}} \\
High school or GED or less & 1.202** & 1.227** & 1.161* & 0.999 & 0.919 \\
Graduate or professional degree & 1.109 & 1.079 & 0.953 & 1.282** & 1.345** \\ 
\multicolumn{6}{l}{\textbf{Race\textsuperscript{5}}} \\
Asian & 1.177 & 1.155 & 1.049 & 1.688*** & 1.256 \\
Black & 1.187 & 1.62*** & 0.972 & 1.315* & 0.853 \\
Other or mixed & 1.224 & 1.23* & 1.152 & 1.074 & 1.005 \\ 
\multicolumn{6}{l}{\textbf{Purchase frequency\textsuperscript{6}}} \\
Less than 5 times per month & 0.91 & 0.972 & 0.818** & 0.983 & 1.004 \\
More than 10 times per month & 1.054 & 1.083 & 1.008 & 1.19 & 0.977 \\ 
\hline
N\textsuperscript{7} & 5820 & 5750 & 4525 & 4390 & 4910 \\
pseudo R-squared & 0.064 & 0.06 & 0.047 & 0.059 & 0.059 \\ \hline
\multicolumn{6}{p{\textwidth}}{\textsuperscript{1}Reference: Male; \textsuperscript{2}Reference: 35-54 years; \textsuperscript{3}Reference: \$50,000 - \$99,999; \textsuperscript{4}Reference: Bachelor's degree \textsuperscript{5}Reference: White; \textsuperscript{6}Reference: 5 to 10 times per month; \textsuperscript{7}Note "I don’t know" responses were excluded from analysis resulting in different values of N.}
\\
\hline
\end{tabular}
\caption{Summary of regression results analyzing the relationships between participant level attributes and survey responses about views on data usage.}
\label{tab:data_use_regression_summary}
\end{table}

We analyze relationships between participant level attributes and survey responses with a logistic regression, estimated separately for each question:
\begin{equation}
\text{{response}} = \text{{share}} + \vec{\text{attributes}}
\label{eq:views_on_data_use_regression}
\end{equation}
The response is a binary variable, where 0 signifies a "No" and 1 signifies a "Yes" (or "Yes, if I/consumers get part of the profit"). 
We excluded "I don't know" responses from this analysis. 
``attributes'' represents a vector of participant characteristics, which were used in Equation~\ref{eq:model_1} and Equation \ref{eq:model_2} and shown in Table~\ref{tab:participant_attributes}.
The model also includes a boolean, ``share'', indicating whether the participant agreed to share their data earlier in the survey. 
This is included because these survey questions followed the decision on data sharing, which could influence responses. Furthermore, this approach enabled us to differentiate the impact of the included attributes from the correlation with sharing behavior identified by the previous model (Equation~\ref{eq:model_1}).

The regression results are presented in Table~\ref{tab:data_use_regression_summary}. For brevity, only odds ratios are shown.
To assess the potential influence of the survey's earlier sections on participants' responses, we performed an additional robustness check by adding terms for the incentive arm and the potential interaction effect of the incentive with data sharing to Equation~\ref{eq:views_on_data_use_regression}. Further details and results of this robustness check are provided in Appendix~\ref{appendix:data_use_opinions_robustness_checks}. We find no significant effects of the incentives or interaction effects between the incentives and data sharing.

Overall a positive response to each question correlated strongly to participants' choice to share data earlier in our study.
Gender differences played a significant role in the survey responses, adding nuance to the previous analyses.
The experiment results (Section~\ref{sec:experiment_results}) showed female participants were
Significantly more likely to share data. 
They were also significantly more likely to respond 'Yes' in support of data access for small businesses, supporting the notion of their greater receptivity to data sharing. 
However, female participants were less likely to support the selling of consumer data by companies and the usage of purchase data by the Census Bureau, suggesting that their willingness to share data does not stem from a disregard for privacy. 

In contrast to the experiment results, age has an important impact on participants’ perspectives on data use.
Experiment results showed younger participants were less likely to share data (with no significant difference for older participants relative to the reference group).
Yet younger participants were more likely to respond "Yes" to all survey questions regarding data use, and older participants were less likely.

For education level, participants with the lowest education levels were more likely to respond in support of companies selling and using consumer data (Q1, Q2, Q3), while participants with the highest education levels were more likely to respond in support of both the Census Bureau and researchers using consumer data (Q4 and Q5).

Regarding the discrepancy found between the responses supporting data use by the Census Bureau (Q4) versus researchers (Q5), we also observe differences across education level, income level, and race.
For the Census Bureau question, Asian and Black participants, and the highest income group (over \$100k), were more likely to respond with 'Yes'. 
Both the highest income group and participants with graduate degrees were also more likely to agree that researchers should be able to use purchase data.
We note these differences between education and income groups for Q5 may be partly due to the participant recruitment, where it's possible that contributing to research is what motivates participants with higher income and education levels to participate, versus supplementing household income.

\section{Discussion}
\label{sec:discussion}

Transacting in digital economies and interacting with online systems has become increasingly unavoidable in everyday life. As major online platforms continue to grow, it is important that researchers and advocates examine their impact on users and society at large. To do this, recent studies have relied heavily on data generated and donated by users of platforms like YouTube and Facebook~\cite{ricksDoesThisButton2022}. As our survey results show, crowdsourcing participants approve of researchers using their contributed data far more than government agencies or corporate actors. Maintaining participant trust in researcher use of crowdsourced data will be critical to supporting future data gathering efforts and platform studies. To do this, researchers must ensure that their crowdsourcing efforts are not done at the expense of user agency and privacy. This section discusses and contextualizes our results with respect to these goals. 

\subsection{Transparency, framing, and improving share rates}

A major issue that researchers face when attempting to crowdsource user data is how to ethically increase participation when asking users to share their data for research. The prevailing approach to solve this problem is to offer cash incentives to participants. This approach is costly and raises ethical concerns: is offering monetary incentives for personal data a coercive practice when a study population includes low-income communities? Our findings suggest that there are other more ethical and less expensive ways to increase data sharing rates.

One way to increase share rates may be to engage in data transparency: show participants their data before asking them to share. Framing data sharing to participants as an altruistic act and showing participants their data was nearly as effective (1.4x) at increasing share rates as offering an additional \$0.50 without data transparency (1.5x).
Meanwhile, offering a lower amount of \$0.20 and showing participants their data was also just as effective (1.5x). For those seeking to simply maximize share rates, the effect of data transparency works even with larger monetary incentives: participants offered \$0.50 were 80\% more likely to share if they were shown their data first (odds ratio 1.49 vs 2.87). These results provide promising evidence that transparency can be a cost-effective way to increase data sharing rates in crowdsourced platform data collection projects. 

Our survey also suggests benefits of highlighting how crowdsourced data will be used for research when asking participants to opt-in to sharing, given that a majority of participants approved of researcher use of their data. Compared to use by businesses or governments, participants viewed researchers using their purchase history data very favorably, with an approval rate of 58\% for researcher use compared to only 23\% for government census use. That participants were more willing to share their data for research underscores the important role that researchers can play in holding powerful players in the online platform space accountable. Researchers, particularly those in the HCI and CSCW communities, are in a unique position to crowdsource data to reveal potential issues and bring them to the attention of advocates and regulators. However, this important role makes using ethical and transparent practices (which conveniently can increase data sharing rates) all the more important in order to maintain the apparent trust that researchers currently enjoy.

\subsection{Relationships between share rates, gender, age, and education}

One important finding from our study is that certain demographic groups either have different baseline share rates or view how their data should be used quite differently. Gender was a major factor—women were significantly more likely to share their data than men. Yet the final survey questions about perceptions on data use added some nuance to this difference. These results showed female participants disapproved of third parties using purchases data (except in the case of small businesses competing with larger companies) when compared to their male counterparts, suggesting that their willingness to share data does not stem from a disregard for privacy. 
Researchers working with participants who come from less-educated backgrounds should be particularly careful. Participants with a high school degree or less were significantly more likely to share their data, controlling for other factors. Although we did not study this specifically, less-educated participants may lack the tools to fully evaluate the potential risks of sharing their personal data.

Some of our results also suggest interesting lines of future research related to demographics and attitudes towards data sharing and use. While younger participants approved uses of their data by various parties in our data use survey, they were less likely to share their data. Why is this the case? Younger participants who are more de-sensitized to a quantified and instrumented society may feel less skeptical of government or researcher use of personal data, but may also be more aware of what their data is able to reveal about them. This could mean that in the future, crowdsourced collection efforts should focus on the governance of how the resulting data is used in order to increase participation from younger populations.

\subsection{The privacy paradox and effectiveness of cash incentives}
Evidence that supports the privacy paradox suggests that people claim to value privacy more than they do in practice~\cite{athey2017digital}. 
This has important implications for researchers seeking to crowdsource data from users of online platforms or study data sharing behaviors for a number of reasons. To our knowledge, this is the first large empirical study to test and contrast both real and hypothetical data sharing scenarios. Our results provide some evidence of the privacy paradox with two nuanced findings that can help inform future crowdsourcing efforts.

First, our results show the effect of the privacy paradox is maintained regardless of monetary incentives and data transparency. Data transparency increases share rates in both real and hypothetical scenarios.
Furthermore,  in both real and hypothetical scenarios, higher monetary incentives correlate with higher share rates. We find a linear relationship between the value of cash incentives offered to participants and share rates in both conditions. Consistent with the privacy paradox, this linear effect, shown in Figure~\ref{fig:real_vs_hypothetical}~(left), is diminished in the hypothetical condition: cash incentives increase hypothetical share rates, but less effectively than in the real condition. 
These consistent relationships between data transparency and increased cash incentives found across the real and hypothetical sharing scenarios suggest that despite the privacy paradox, hypothetical scenarios might still be useful for studying data sharing behaviors. 
If these relationships hold in other sharing contexts, then hypothetical data sharing behavior measured through surveys could be used to approximate real-world behavior. 

\subsection{Monetary incentives and limitations of the experiment context}
Our experiment tested a limited range of real bonus payments (\$0.05, \$0.20, \$0.50) on participants' likelihood to share data. 
This presents a limitation in regards to generalizability. Future work should test the robustness of the linear relationship we see between incentive value and share rates when wider ranges of monetary incentives are used. We encourage other researchers to explore the impact of monetary incentives, especially in combination with transparency, more deeply in future work.

However, using monetary incentives may unduly influence low-income participants to share data, particularly on low-paid crowdwork platforms. We chose to limit our bonuses to \$0.50 and below in order to mitigate this risk. 
When we tested for interaction effects between income group and bonus amounts, we did not find evidence that bonus amounts disproportionately change share rates among low-income participants in our study. 
This suggests that with small bonus amounts, this risk may be minimal. However, given how small our bonuses were, this may not be the case when larger monetary incentives are offered and care should be taken by future researchers offering monetary incentives in exchange for personal data.

Given the experiment context, our results do not imply a monetary amount by which people value their data or privacy.  Instead they show how monetary rewards influenced data sharing in an IRB-approved study at a well-known and reputable university.
Furthermore, this context likely increased participants’ trust, impacted their willingness to share data, and possibly lowered the amount for which they were willing to accept in exchange for the data.
Further work must be done to test the generalizability of our results, such as by conducting experiments outside the context of a study by a well known and reputable institution.

\subsection{Limitations in generalizability and future work} 
This study is limited to the 5 incentives and 2 transparency treatments we experimented with. Further research that expands on and experiments with other concepts of “data transparency” and various kinds of incentives are also needed. For example, how might providing participants with a promise of “insight” into their behavior impact share rates? 

Further research must also be done to test how our findings generalize in other data sharing contexts. We also only examined data sharing within the specific context of an online survey and with a population of crowd workers. 
Furthermore, our experiment only tested participants’ likelihood of sharing data given that they had already downloaded it from Amazon, the platform we study here. It could be argued that the process of downloading, then sharing, the data from Amazon is a more intensive and therefore more difficult task to incentivize than simply opting in to share. Researchers asking users “in the wild” to contribute their data from a platform will also face the additional challenge of asking users to access or export their data. Future work should study how different incentives and framings impact the likelihood of users accessing their data, not just choosing to share it. Other work might test for similar dynamics when users are asked to share data collected through other tools, such as browser extensions. 

Another context-specific factor that may impact data sharing is the type of data collected and the platform from which data are collected. 
In our study, we collected Amazon purchase histories data because we are interested in the future analyses this dataset will enable and given we had not seen other researchers develop a publicly available dataset of purchase histories, we believed this would be a contribution to the research community.
We determined which data fields were of interest and our data collection process was careful to drop all other fields. In particular, we avoided collecting any address or payment data and other explicitly personal identifiable information (PII), and made this clear to study participants. Our results regarding the impact of framing and incentives may not extend to more sensitive types of data, such as contact information or PII.
Furthermore, more work must be done to test how our findings generalize beyond the context of collecting Amazon data.
Consider recent work that has investigated people's willingness to donate different types of data from different types of platforms~\cite{pfiffner2023leveraging}. The researchers found a high willingness to donate YouTube data compared to Facebook, Instagram, or Google, and investigated reasons why. However, we note that their study only posed participants with a hypothetical question about data sharing, versus a real data sharing context such as in our work. 
Future research can build on both their findings and the present work to study how framing and incentives have different impacts on users' willingness to share data across platforms and types of data, and in a real sharing context.

A prerequisite for future work in crowdsourcing, as described above, is users' ability to obtain their data from platforms. Fortunately, privacy regulation around the world have required companies to offer features enabling consumers to access their data (for example the EU~\cite{gdpr2016}, California~\cite{ccpa2018}, Brazil~\cite{lgpd2018}, and Singapore~\cite{pdpa2012} all give consumers data access rights). The work presented here focuses on the U.S. context where there is no general national right to data access. This limitation was highlighted in our own data collection process when the feature our data collection tool relied on was taken offline in March 2023~\footnote{The data export tool used in this study was accessible until March 2023 at \url{https://www.amazon.com/gp/b2b/reports}. Amazon provides a newer tool to satisfy data requests, available at \url{https://www.amazon.com/hz/privacy-central/data-requests/preview.html}. However, obtaining data this way takes a variable number of days~\cite{mazurov2022}, which adds friction to data crowdsourcing.}. 
Recently, a group of data privacy researchers brought the importance of data access to facilitate research to the attention of the U.S. Federal Trade Commission (FTC)~\cite{berke2022comment}. Their comment argued for the potential benefits of giving consumers access to their data to share it with researchers and consumer advocacy groups, who are well positioned to surface potential harms stemming from the corporate use of this data. We believe  future work in crowdsourcing, as well as downstream applications of the Amazon data we publish, can underscore the importance of enabling platform users easy access to their own data to then share, with informed consent, with researchers.

\section{Conclusion}
\label{sec:conclusion}
Datasets generated by users are becoming critical tools in algorithmic auditing, studying human behavior, and uncovering digital inequities. 
This makes ethically collecting user-generated data while maintaining privacy and respecting consent increasingly important. In this paper, we aim to provide researchers with information and design patterns they can use to both more ethically and effectively crowdsource datasets. 
To do this, we use an innovative approach to crowdsourcing user data to study privacy perceptions and data sharing behaviors with a sample of over 6,000 users. 
We ask participants to download and then opt-in to share their Amazon purchase history data with us, and test how different monetary incentives and levels of data transparency influence how often they agree to share their data. 

This paper offers four major contributions.
First, we share the design of our tool as an example of how to crowdsource user data in a way that prioritizes participant consent, an important addition to the ongoing literature on collecting user data. 
Second, our results provide guidance to future data crowdsourcing efforts by quantifying the positive effect of data transparency, the added impact of monetary incentives, and the importance of participant demographics, on sharing rates. 
Third, we provide nuanced evidence for the privacy paradox in data sharing. Our unique empirical study demonstrates that cash incentives and data transparency both impact the likelihood that participants hypothetically share their data, but where the impact is smaller than in real transactions. 
Finally, we show that among our study participants, crowdsourced data use by researchers is a largely supported practice, with a higher overall rate of approval than government (census) use or private corporate data use. 
We hope these findings can improve the norms and efficacy in future efforts crowdsourcing user-generated data in order to benefit future research.

\section*{Data and Code Availability}
All code and data described in this paper are available in an open source repository:\\
\url{https://github.com/aberke/amazon-study}.\\

This includes all code that enabled the experiment embedded in the survey. It also includes all preprocessing code and data analyses, which provides further details for the analyses described in the paper. 
The survey data (both prescreen and main study) are also made available in the repository. Raw data are not provided, as they contain PII and other fields stripped to conform with IRB guidelines - the data provided represents data after preprocessing.

The Amazon purchase histories collected through the process described in this paper are also available and further described in~\cite{berke2024a}.

\begin{acks}
The authors thank all of the study participants - those who chose to share their data as well as those who did not - for their contributions. 
\end{acks}

\bibliographystyle{ACM-Reference-Format}
\bibliography{references}

\appendix
\newpage

\section{Survey tool}
\label{SI:survey_tool}


\subsection{Prescreen survey}
\label{SI:prescreen}

The github repository contains a public preview of the prescreen survey as well as a static written copy of the survey text. The preview version shows the same interactive interface shown to participants. The text version includes all survey questions and response options shown.

See: \url{https://github.com/aberke/amazon-study/tree/master/instrument/prescreen}

The prescreen survey was used to verify potential participants' eligibility for the main study with a series of questions. It also contains an attention check.
If participants provided an answer indicating they were ineligible or failed the attention check, they were immediately sent to the end of the survey to save participant time.
Participants who passed the prescreen were invited to the main study.

\subsection{Main study survey}
\label{SI:survey}

The github repository contains a public preview of the main survey. It also contains a written copy of the text, survey questions, and response options shown to participants.

See: \url{https://github.com/aberke/amazon-study/tree/master/instrument}.

\begin{figure}[htbp]
    \begin{adjustwidth}{-0.7cm}{}
\centering
\includegraphics[width=0.52\textwidth]{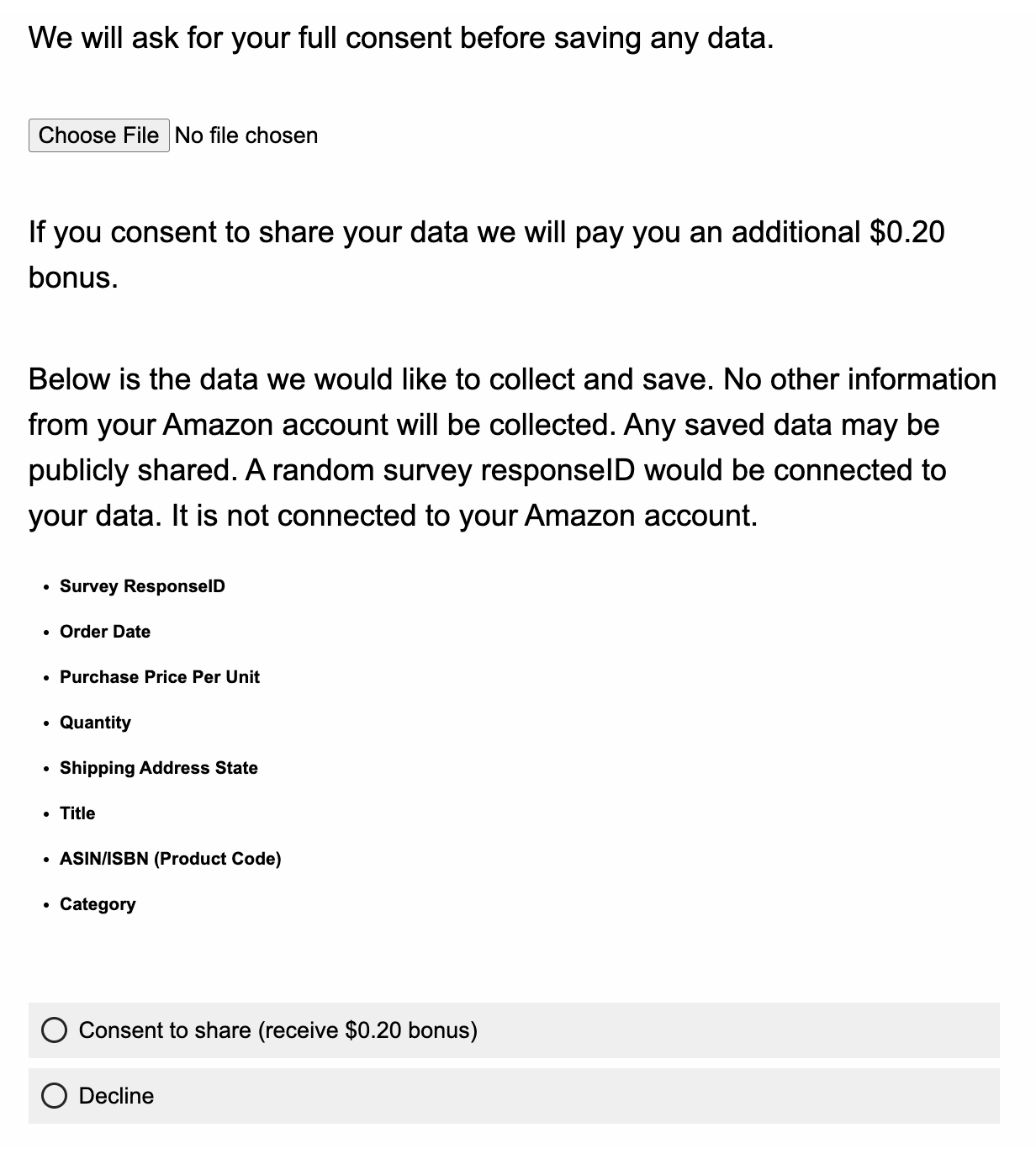}
\includegraphics[width=0.52\textwidth]{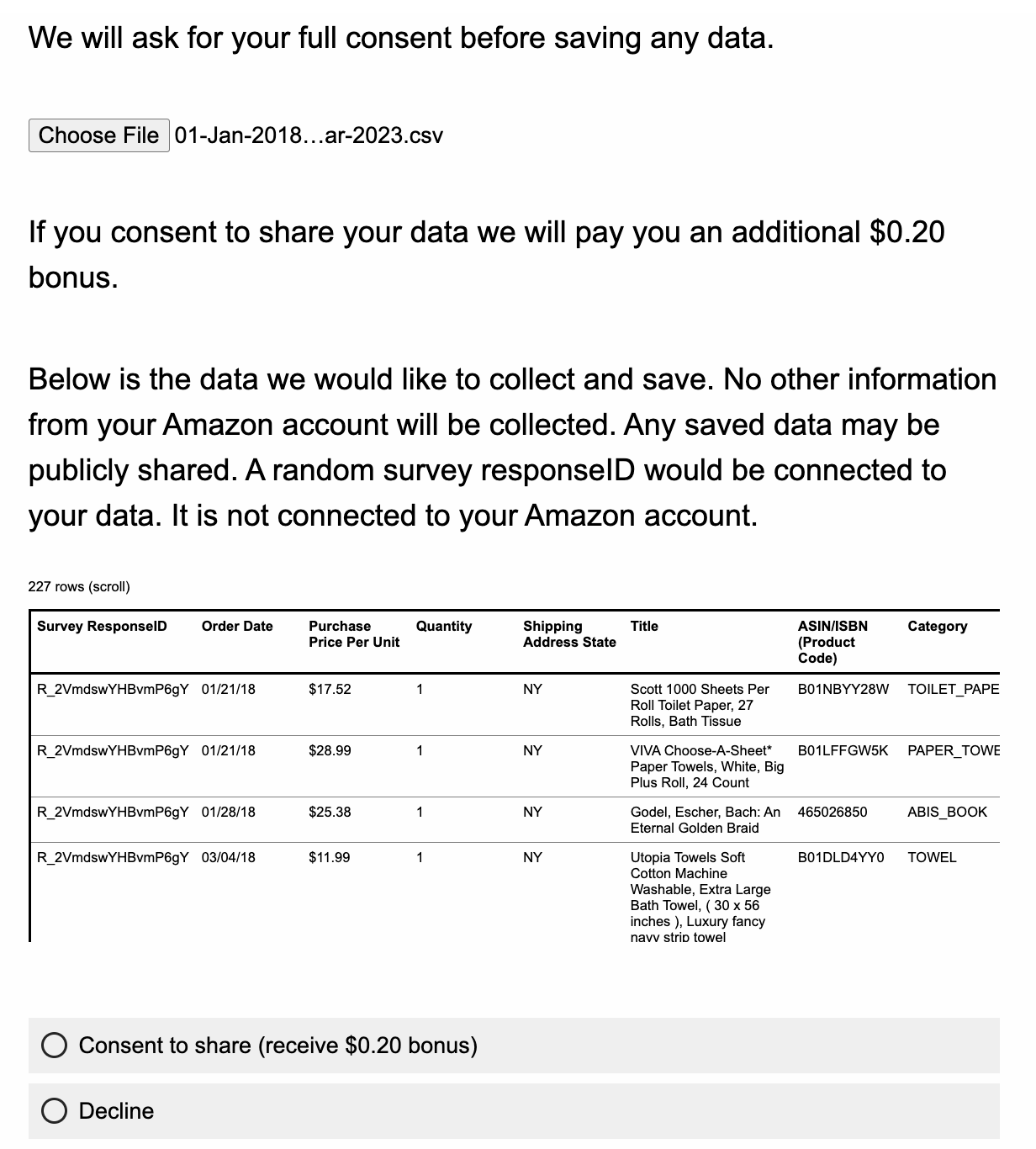}
\end{adjustwidth}
\caption{Screenshots from the survey's share request section. The experiment had a 2x5 factorial design, with 2 "transparency" treatments and 5 "incentives", with 10 total experiment arms. Shown is the experiment arm with the "transparent" and "\$0.20 bonus" treatments. Left: Interface before inserting Amazon data file. Right: Interface after inserting Amazon data file. Software within the browser stripped the data file to only include the data columns the survey text described to participants.
No data left participant machines unless they clicked "Consent to share". 
The interface for the "transparent" treatment presented participants with all rows and columns of data that would be collected, within a scrollable interface, before they chose to consent or decline to share. The "non-transparent" treatment only showed the data columns.}
\label{fig:share_prompt_example_20_showdata}
\end{figure}

Figure~\ref{fig:share_prompt_example_20_showdata} shows an example of the interface participants saw.
It shows screenshots from the survey's share request section for the experiment arm with the "transparent" and "\$0.20 bonus" treatments.
Below we provide the text for the 5 different incentive treatments. 

\vspace{0.5cm}

\begin{mdframed}

\noindent Click to insert the file from Amazon below. 

\noindent We will ask for your full consent before saving any data. 
\vspace{0.2cm}

\noindent [file upload button]

\noindent [Note: Error text is shown upon validation failures]
\vspace{0.2cm}

\noindent [Additional incentive language specific to experiment arm]
\vspace{0.2cm}

\noindent Below is the data we would like to collect and save. No other information from your Amazon account will be collected. Any saved data may be publicly shared.
A random survey responseID would be connected to your data. It is not connected to your Amazon account. 
[Note: The below columns are replaced with the table view for the transparent condition]

\begin{itemize}
    \item \textbf{Survey ResponseID}
    \item \textbf{Order Date}
    \item \textbf{Purchase Price Per Unit}
    \item \textbf{Quantity}
    \item \textbf{Shipping Address State}
    \item \textbf{Title}
    \item \textbf{ASIN/ISBN (Product Code)}
    \item \textbf{Category}
\end{itemize}

\noindent [button: Consent to share]
\newline
\noindent [button: Decline]
\end{mdframed}

\vspace{0.5cm}

Below provides the additional incentive language specific to experiment arms.

\begin{mdframed}

\noindent \textbf{Control}: No additional language.

\vspace{0.3cm}
\noindent \textbf{Altruism}: \textit{Why are we asking you to share?}

\noindent We are crowdsourcing data to democratize access to it as a public good and we are asking for your help.

\noindent Large amounts of data from people like you are valuable!

\noindent Amazon and the companies they transact with already use and profit from your data. 

\noindent Your data can also benefit researchers, organizations, and communities trying to help people via the knowledge your data can provide.
\vspace{0.3cm}

\noindent \textbf{\$X bonus (X=0.05, 0.20, 0.50):}

\noindent If you consent to share your data we will pay you an additional \$X bonus.

\end{mdframed}

\newpage
\section{Survey data detailed}
\label{SI:detailed_survey_data}


Here we provide more detailed breakdowns of survey participants.
Table \ref{tab:participant_attributes_detailed} shows the detailed participant attributes in the fine grained categories that were collected in the survey. Unlike Table \ref{tab:participant_attributes} in the main text, the survey \% in this table includes answers that do not correspond to census data categories (e.g. "Prefer not to say").

Table \ref{tab:participant_state} shows the number of survey participants from each U.S. state as well as Washington DC and Puerto Rico, in comparison to population estimates from the U.S. Census Bureau \cite{census2022SCPRC-EST2022-18+POP}.
Census data are for the 18+ population in order to provide a better comparison to the survey participants, who were required to be 18+.

\begin{table}[]
\centering
\small
\begin{tabular}{lccc}
\hline
& \multicolumn{2}{c}{\textbf{Survey}} & \textbf{Census} \\
\hline
\textbf{Attribute} & \textbf{N} & \textbf{\%} & \textbf{\%} \\ 
\hline
\textbf{Gender} & & & \\
Female & 3132 & 49.5\% & 51.0\% \\
Male & 3020 & 47.7\% & 49.0\% \\
Other & 149 & 2.4\% & - \\
Prefer not to say & 24 & 0.4\% & - \\ 
\textbf{Age} & & & \\
18 - 24 years & 990 & 15.7\% & 12.0\% \\
25 - 34 years & 2312 & 36.6\% & 17.4\% \\
35 - 44 years & 1545 & 24.4\% & 16.8\% \\
45 - 54 years & 824 & 13.0\% & 15.5\% \\
55 - 64 years & 461 & 7.3\% & 16.1\% \\
65 and older & 193 & 3.1\% & 22.2\% \\
\textbf{Household income} & & & \\
Less than \$25,000 & 855 & 13.5\% & 17.1\% \\
\$25,000 - \$49,999 & 1443 & 22.8\% & 18.4\% \\
\$50,000 - \$74,999 & 1341 & 21.2\% & 18.6\% \\
\$75,000 - \$99,999 & 974 & 15.4\% & 11.7\% \\
\$100,000 - \$149,999 & 977 & 15.4\% & 14.6\% \\
\$150,000 or more & 605 & 9.6\% & 19.5\% \\
Prefer not to say & 130 & 2.1\% & - \\
\textbf{Education level} & & & \\
Some high school or less & 60 & 0.9\% & 9.6\% \\
High school diploma or GED & 2280 & 36.0\% & 55.6\% \\
Bachelor's degree & 2834 & 44.8\% & 22.1\% \\
Graduate or professional degree & 1086 & 17.2\% & 12.7\% \\
Prefer not to say & 65 & 1.0\% & - \\
\textbf{Race} & & & \\
White & 4825 & 76.3\% & 75.8\% \\
Asian & 551 & 8.7\% & 6.1\% \\
Black or African American & 440 & 7.0\% & 13.6\% \\
American Indian and Alaska Native & 38 & 0.6\% & 1.3\% \\
Native Hawaiian and Other Pacific Islander & 7 & 0.1\% & 0.3\% \\
Two or More Races & 330 & 5.2\% & 2.9\% \\
Other & 134 & 2.1\% & - \\
\textbf{Online purchase frequency} & & & \\
Less than 5 times per month & 4081 & 64.5\% & - \\
5 - 10 times per month & 1760 & 27.8\% & - \\
More than 10 times per month & 484 & 7.7\% & - \\ 
\hline
\end{tabular}
\caption{Survey participant attributes as they were collected, compared to census estimates.
}
\label{tab:participant_attributes_detailed}
\end{table}

\begin{table}[]
\centering
\footnotesize
\begin{tabular}{lccccc}
\hline
& \multicolumn{2}{c}{\textbf{Census}} & \multicolumn{2}{c}{\textbf{Survey}} \\
\hline
\textbf{US state/territory} & \textbf{N} & \textbf{\%} & \textbf{N} & \textbf{\%} \\ \hline
Alabama & 3962734 & 1.5\% & 82 & 1.3\% \\
Alaska & 557060 & 0.2\% & 11 & 0.2\% \\
Arizona & 5770187 & 2.2\% & 129 & 2.0\% \\
Arkansas & 2348518 & 0.9\% & 55 & 0.9\% \\
California & 30523315 & 11.6\% & 662 & 10.5\% \\
Colorado & 4624351 & 1.8\% & 109 & 1.7\% \\
Connecticut & 2895175 & 1.1\% & 56 & 0.9\% \\
Delaware & 810269 & 0.3\% & 14 & 0.2\% \\
District of Columbia & 547328 & 0.2\% & 16 & 0.3\% \\
Florida & 17948469 & 6.8\% & 416 & 6.6\% \\
Georgia & 8402753 & 3.2\% & 200 & 3.2\% \\
Hawaii & 1142870 & 0.4\% & 21 & 0.3\% \\
Idaho & 1475629 & 0.6\% & 20 & 0.3\% \\
Illinois & 9861901 & 3.7\% & 290 & 4.6\% \\
Indiana & 5263114 & 2.0\% & 160 & 2.5\% \\
Iowa & 2476028 & 0.9\% & 55 & 0.9\% \\
Kansas & 2246318 & 0.9\% & 52 & 0.8\% \\
Kentucky & 3507735 & 1.3\% & 111 & 1.8\% \\
Louisiana & 3528548 & 1.3\% & 70 & 1.1\% \\
Maine & 1137442 & 0.4\% & 20 & 0.3\% \\
Maryland & 4818071 & 1.8\% & 129 & 2.0\% \\
Massachusetts & 5644540 & 2.1\% & 151 & 2.4\% \\
Michigan & 7924418 & 3.0\% & 211 & 3.3\% \\
Minnesota & 4423022 & 1.7\% & 118 & 1.9\% \\
Mississippi & 2261996 & 0.9\% & 38 & 0.6\% \\
Missouri & 4813049 & 1.8\% & 97 & 1.5\% \\
Montana & 889114 & 0.3\% & 9 & 0.1\% \\
Nebraska & 1491246 & 0.6\% & 42 & 0.7\% \\
Nevada & 2487994 & 0.9\% & 65 & 1.0\% \\
New Hampshire & 1142307 & 0.4\% & 25 & 0.4\% \\
New Jersey & 7267590 & 2.8\% & 167 & 2.6\% \\
New Mexico & 1653831 & 0.6\% & 33 & 0.5\% \\
New York & 15687863 & 6.0\% & 401 & 6.3\% \\
North Carolina & 8404094 & 3.2\% & 217 & 3.4\% \\
North Dakota & 596486 & 0.2\% & 6 & 0.1\% \\
Ohio & 9193508 & 3.5\% & 267 & 4.2\% \\
Oklahoma & 3066654 & 1.2\% & 70 & 1.1\% \\
Oregon & 3403149 & 1.3\% & 114 & 1.8\% \\
Pennsylvania & 10347543 & 3.9\% & 342 & 5.4\% \\
Rhode Island & 889822 & 0.3\% & 22 & 0.3\% \\
South Carolina & 4164762 & 1.6\% & 81 & 1.3\% \\
South Dakota & 690659 & 0.3\% & 11 & 0.2\% \\
Tennessee & 5513202 & 2.1\% & 131 & 2.1\% \\
Texas & 22573234 & 8.6\% & 476 & 7.5\% \\
Utah & 2449192 & 0.9\% & 49 & 0.8\% \\
Vermont & 532307 & 0.2\% & 16 & 0.3\% \\
Virginia & 6816709 & 2.6\% & 183 & 2.9\% \\
Washington & 6139213 & 2.3\% & 154 & 2.4\% \\
West Virginia & 1423234 & 0.5\% & 27 & 0.4\% \\
Wisconsin & 4646910 & 1.8\% & 116 & 1.8\% \\
Wyoming & 451267 & 0.2\% & 6 & 0.1\% \\
Puerto Rico & 2703450 & 1.0\% & 0 & 0.0\% \\ \hline
\end{tabular}
\caption{Participation by US state/territory compared to population estimates from the US census.}
\label{tab:participant_state}
\end{table}

\clearpage
\newpage
\section{Additional analyses and tables}
\label{SI:additional_analyses_and_tables}

\subsection{Analysis for non-response bias}
\label{SI:nonresponse_bias}

The last question of the prescreen asked participants whether they were interested in the main study, given that it would ask them to log into their Amazon account. Only participants who answered this question were invited to participate in the main study.
We know this question presented a potential privacy concern for some participants.

We analyze whether there were demographic differences in the response to this question that may have contributed to non-response bias in the main study.

To assess this we use the following logistic regression:
\begin{equation}
\text{opt\_out} = \text{age} + \text{sex} + \text{race}
\label{eq:opt_out_analysis}
\end{equation}						
The variable `opt\_out` defined as 1 if answered not interested in participating, 0 otherwise.

Note this analysis is only possible for participants recruited via Prolific who supplied their demographic information to Prolific, and the variables are hence defined by Prolific.
Prolific provides an ‘Ethnicity variable’. We collapse the categories to match the race groups used in our main study analysis: White, Black Asian, Other or mixed

To conduct the analysis we first restrict the data to participants whose following demographic variables are available: Age, Sex , Race (where we collapsed the Ethnicity variable).
There were a total of 13,154 participants in the model, with an opt out rate of 0.102.
Details are in Table~\ref{tab:response_bias_regression}.
Results show older participants were significantly more likely to opt out (p<0.05).
This exacerbated the bias in demographics (older population are under-represented in our sample). It also means there was a non-response bias that should be taken into account when interpreting the experiment results with respect to age. In particular, results show younger participants were significantly less likely to share. We must consider that older participants worried about privacy may have already opted out.

\begin{table}[h]
\centering
\small
\begin{tabular}{lccc}
\hline
\textbf{Predictor} & \textbf{B (log odds)} & \textbf{Odds ratio} & \textbf{95\% CI for Odds Ratio} \\ \hline
Intercept & -2.153*** & 0.116 & [0.102, 0.133] \\
\textbf{Age (Reference: 35-44 years)} & & & \\
18-24 years & -0.107 & 0.898 & [0.745, 1.083] \\
25-34 years & -0.074 & 0.929 & [0.796, 1.085] \\
45-54 years & 0.184 & 1.201 & [0.988, 1.462] \\
55-64 years & 0.233* & 1.263 & [1.008, 1.582] \\
65 and older & 0.539*** & 1.714 & [1.294, 2.272] \\
\textbf{Gender (Reference: Male)} & & & \\
Female & -0.063 & 0.939 & [0.837, 1.053] \\
\textbf{Race (Reference: White)} & & & \\
Asian & -0.141 & 0.868 & [0.690, 1.091] \\
Black & -0.057 & 0.945 & [0.753, 1.185] \\
Other or mixed & -0.080 & 0.924 & [0.757, 1.126] \\
\hline
\multicolumn{4}{l}{N = 13154} \\
\multicolumn{4}{l}{Pseudo R-squared = 0.004} \\
\hline
\end{tabular}
\caption{Results for logistic regression assessing whether opting out of the main study was related to demographics (equation~\ref{eq:opt_out_analysis}).
\newline
Note: *p<0.05; **p<0.01; ***p<0.001.}
\label{tab:response_bias_regression}
\end{table}

\clearpage
\subsection{Verifying even distribution across experiment arms with a chi-square test of homogeneity}
\label{SI:attrition_bias}

A potential source of response bias in the survey design is that participants in specific experiment arms will be less likely to complete the survey due to how their data and share incentive are presented to them.
Given participants were randomly assigned to their experiment arm at the start of the survey at equal rates, we should expect the number of completions across experiment arms to match in the absence of such bias.

We use a chi-square test of homogeneity to test for such response bias by checking whether the number of completions is consistent across the 10 experiment arms.
Specifically, we test the following null hypothesis with a significance level of 0.05: Completions count is the same across experimental arms.
$X^2=3.953$, $p=.914$. We do not reject the null hypothesis.

\subsection{Additional tables for data sharing results
}
\label{SI:additional_tables_data_sharing}

See Table \ref{tab:overall_share_rates} for the number of participants and share rates for each experiment arm.

\begin{table}[h]
\centering
\small
\begin{tabular}{lcc}
\hline
& \textbf{N} & \textbf{Share Rate} \\ 
\hline
\textbf{Non-transparent} & 3208 & \\
\hline
Control & 636 & 0.763 \\
\$0.05 & 647 & 0.720 \\
\$0.20 & 648 & 0.765 \\
\$0.50 & 624 & 0.830 \\
Altruism & 653 & 0.767 \\
\hline
\textbf{Transparent} & 3117 & \\
\hline
Control & 639 & 0.779 \\
\$0.05 & 618 & 0.785 \\
\$0.20 & 602 & 0.834 \\
\$0.50 & 616 & 0.888 \\
Altruism & 642 & 0.816 \\ \hline
\end{tabular}
\caption{Participants and share rates by experiment arm.}
\label{tab:overall_share_rates}
\end{table}

\begin{table}[]
\centering
\small
\begin{tabular}{lcccc}
\hline
\textbf{Predictor} & \textbf{B (log odds)} & \textbf{Odds ratio} & \textbf{95\% CI for Odds Ratio} \\ \hline
Intercept & 1.0*** & 2.717 & [2.050, 3.602] \\
Transparent & 0.374*** & 1.454 & [1.277, 1.656] \\
\multicolumn{4}{l}{\textbf{Incentive (Reference: control)}} \\
Altruism & 0.092 & 1.097 & [0.799, 1.505] \\
\$0.05 & -0.186 & 0.830 & [0.607, 1.135] \\
\$0.20 & -0.008 & 0.992 & [0.720, 1.366] \\
\$0.50 & 0.659*** & 1.932 & [1.358, 2.749] \\
\multicolumn{4}{l}{\textbf{Gender (Reference: Male)}} \\
Female & 0.367*** & 1.443 & [1.267, 1.644] \\
\multicolumn{4}{l}{\textbf{Age (Reference: 35 - 54 years)}} \\
18 - 34 years & -0.175* & 0.840 & [0.729, 0.967] \\
55 and older & -0.062 & 0.940 & [0.745, 1.184] \\
\multicolumn{4}{l}{\textbf{Household income (Reference: \$50,000-\$99,999)}} \\
Less than \$50,000 & 0.121 & 1.129 & [0.810, 1.574] \\
\$100,000 or more & 0.065 & 1.067 & [0.753, 1.513] \\
\multicolumn{4}{l}{\textbf{Education (Reference: Bachelor's degree)}} \\
High school or GED or less & 0.16* & 1.174 & [1.011, 1.362] \\
Graduate or professional degree & 0.129 & 1.138 & [0.947, 1.367] \\
\multicolumn{4}{l}{\textbf{Race (Reference: White)}} \\
Asian & -0.56*** & 0.571 & [0.463, 0.705] \\
Black & -0.071 & 0.931 & [0.719, 1.206] \\
Other or mixed & 0.019 & 1.020 & [0.795, 1.308] \\
\multicolumn{4}{l}{\textbf{Purchase frequency (Reference: 5-10 times/month)}} \\
Less than 5 times per month & -0.034 & 0.967 & [0.832, 1.123] \\
More than 10 times per month & -0.18 & 0.835 & [0.647, 1.077] \\
\multicolumn{4}{l}{\textbf{Household income $\times$ Incentive}} \\
Less than \$50,000 $\times$ altruism & -0.025 & 0.976 & [0.614, 1.549] \\
\$100,000 or more $\times$ altruism & -0.168 & 0.845 & [0.517, 1.383] \\
Less than \$50,000  $\times$ \$0.05 & 0.064 & 1.067 & [0.675, 1.685] \\
\$100,000 or more $\times$ \$0.05 & 0.044 & 1.045 & [0.645, 1.692] \\
Less than \$50,000 $\times$ \$0.20 & 0.134 & 1.144 & [0.715, 1.830] \\
\$100,000 or more $\times$ \$0.20 & 0.158 & 1.171 & [0.709, 1.935] \\
Less than \$50,000 $\times$ \$0.50 & -0.084 & 0.919 & [0.549, 1.538] \\
\$100,000 or more $\times$ \$0.50 & -0.177 & 0.838 & [0.483, 1.454] \\ \hline
\multicolumn{4}{l}{N = 5995} \\
\multicolumn{4}{l}{Pseudo R-squared = 0.029} \\
\hline
\end{tabular}
\caption{Results for model testing for interaction effects between incentive and household income.
\newline
$share \sim transparent + incentive + gender + age + household\_income + education + race + purchase\_frequency +  household\_income \times incentive$
\newline
Note: *p<0.05; **p<0.01; ***p<0.001. 
For the interaction between household income and incentive, the reference variables are the same as for the main effects.}
\label{tab:model1_income_incentive_interaction}
\end{table}

\begin{table}[]
\centering
\small
\begin{tabular}{lccc}
\hline
\textbf{Predictor} & \textbf{B (log odds)} & \textbf{Odds Ratio} & \textbf{95\% CI for Odds Ratio} \\ \hline
Intercept & 1.065*** & 2.902 & [2.226, 3.783] \\
\multicolumn{4}{l}{\textbf{Experiment arm (Reference: control)}} \\
Altruism & -0.01 & 0.990 & [0.758, 1.294] \\
\$0.05 & -0.211 & 0.810 & [0.623, 1.053] \\
\$0.20 & -0.007 & 0.993 & [0.758, 1.300] \\
\$0.50 & 0.398** & 1.490 & [1.114, 1.991] \\
Control \& Transparent & 0.223 & 1.250 & [0.945, 1.653] \\
Altruism \& Transparent & 0.318* & 1.374 & [1.036, 1.823] \\
\$0.05 \& Transparent & 0.139 & 1.149 & [0.870, 1.518] \\
\$0.20 \& Transparent & 0.407** & 1.503 & [1.121, 2.014] \\
\$0.50 \& Transparent & 1.053*** & 2.865 & [2.053, 3.999] \\
\multicolumn{4}{l}{\textbf{Gender (Reference: Male)}} \\
Female & 0.368*** & 1.444 & [1.268, 1.645] \\
\multicolumn{4}{l}{\textbf{Age (Reference: 35 - 54 years)}} \\
18 - 34 years & -0.175* & 0.840 & [0.729, 0.967] \\
55 and older & -0.064 & 0.938 & [0.744, 1.183] \\
\multicolumn{4}{l}{\textbf{Household income (Reference: \$50,000-\$99,999)}} \\
Less than \$50,000 & 0.15 & 1.161 & [0.994, 1.357] \\
\$100,000 or more & 0.048 & 1.050 & [0.890, 1.238] \\
\multicolumn{4}{l}{\textbf{Education (Reference: Bachelor's degree)}} \\
High school or GED or less & 0.162* & 1.176 & [1.013, 1.365] \\
Graduate or professional degree & 0.129 & 1.137 & [0.947, 1.366] \\
\multicolumn{4}{l}{\textbf{Race (Reference: White)}} \\
Asian & -0.559*** & 0.572 & [0.463, 0.706] \\
Black & -0.071 & 0.932 & [0.719, 1.207] \\
Other or mixed & 0.019 & 1.019 & [0.794, 1.307] \\
\multicolumn{4}{l}{\textbf{Purchase frequency (Reference: 5-10 times/month)}} \\
Less than 5 times per month & -0.037 & 0.963 & [0.829, 1.119] \\
More than 10 times per month & -0.175 & 0.840 & [0.651, 1.083] \\ \hline
N = 5995 & & & \\
Pseudo R-squared = 0.029 & & & \\
\hline
\end{tabular}
\caption{Full regression results for model 2. Shows effects of the 10 experiment arms: the relative impact of the incentive treatments for both the transparent and non-transparent conditions.
\newline
Note: *p<0.05; **p<0.01; ***p<0.001.
The reference experiment arm is control \& non-transparent data condition. Only "transparent" is included in the table rows and otherwise "non-transparent" is assumed.}
\label{tab:model2_results_full}
\end{table}

\clearpage
\subsection{Real versus hypothetical share incentives}
\label{SI:real_v_hypothetical_share_incentives}

Data for analyzing change in share rates due to real versus hypothetical incentives using all experimental arms are shown in Table~\ref{tab:real_vs_hypothetical_direct} with the 
methodology further detailed in Table~\ref{tab:real_hypo_method_details} and below.

\begin{table}[h]
\centering
\small
\begin{tabular}{lcc}
\hline
& \textbf{Real Share Rate} & \textbf{Hypothetical Share Rate} \\ \hline
\textbf{Non-transparent} &  & N=636\\
Control & 0.763 & - \\
\$0.05 & 0.720 & 0.781 \\
\$0.20 & 0.765 & 0.785 \\
\$0.50 & 0.830 & 0.792 \\
\$1.00 & - & 0.818 \\
Any \$X & - & 0.939 \\
\hline
\multicolumn{3}{l}{\textbf{Least squares (share = a + $\beta$ x amount)}} \\
a & 0.712 & 0.777 \\
b & 0.240 & 0.039 \\
Pearson r & 0.996 & 0.983 \\
\hline
\textbf{Transparent} & &  N=639\\
Control & 0.779 & - \\
\$0.05 & 0.785 & 0.806 \\
\$0.20 & 0.834 & 0.811 \\
\$0.50 & 0.888 & 0.818 \\
\$1.00 & - & 0.834 \\
Any \$X & - & 0.961 \\
\hline
\multicolumn{3}{l}{\textbf{Least squares (share = a + $\beta$ x amount)}} \\
a & 0.780 & 0.805 \\
b & 0.222 & 0.029 \\
Pearson r & 0.987 & 0.998 \\ \hline
\end{tabular}
\caption{Real and hypothetical share rates where the control arm is used to compute hypothetical share rates. Note the coefficients are scaled to \$1 for readability, while our bonus amounts were all less than \$1.}
\label{tab:real_vs_hypothetical_incremental} 
\end{table}

\begin{table}[]
\centering
\small
\begin{tabular}{lccccccc}
\hline
\textbf{Incentive} & \textbf{N} & \multicolumn{1}{p{0.8cm}}{\centering \textbf{Real\\ Shares}} & \multicolumn{1}{p{1.5cm}}{\centering \textbf{Real Share\\ Rate}} & \multicolumn{1}{p{1cm}}{\centering \textbf{Real\\ Change}} & \multicolumn{1}{p{2.2cm}}{\centering \textbf{Additional Hypothetical Shares}} & \multicolumn{1}{p{2cm}}{\centering \textbf{Hypothetical\\ Share Rate}} & \multicolumn{1}{p{1.5cm}}{\centering \textbf{Hypothetical\\ Change}} \\ \hline
\multicolumn{8}{l}{\textbf{Non-transparent}} \\
\hspace{1em} Control & 636 & 485 & 0.763 & - & - & - & - \\
\hspace{1em} \$0.05 & 647 & 466 & 0.720 & -0.043 & 12 & 0.781 & 0.019 \\
\hspace{1em} \$0.20 & 648 & 496 & 0.765 & 0.045 & 17 & 0.747 & 0.026 \\
\hspace{1em} \$0.50 & 624 & 518 & 0.830 & 0.065 & 27 & 0.807 & 0.042 \\
\hspace{1em} \$1.00 & - & - & - & - & 18 & 0.859 & 0.029 \\
\multicolumn{8}{l}{\textbf{Transparent}} \\
\hspace{1em} Control & 639 & 498 & 0.779 & - & - & - & - \\
\hspace{1em} \$0.05 & 618 & 485 & 0.785 & 0.006 & 17 & 0.806 & 0.027 \\
\hspace{1em} \$0.20 & 602 & 502 & 0.834 & 0.049 & 21 & 0.819 & 0.034 \\
\hspace{1em} \$0.50 & 616 & 547 & 0.888 & 0.054 & 22 & 0.870 & 0.037 \\
\hspace{1em} \$1.00 & - & - & - & - & 18 & 0.917 & 0.029 \\ \hline
\end{tabular}
\caption{Data for analyzing change in share rates due to real versus hypothetical incentives using all experimental arms}
\label{tab:real_vs_hypothetical_direct}
\end{table}

\begin{table}[]
\centering
\small
\begin{tabular}{lccccccc}
\hline
\textbf{Incentive} & \textbf{n} & \multicolumn{1}{p{0.8cm}}{\centering \textbf{Real\\ Shares}} & \multicolumn{1}{p{1.5cm}}{\centering \textbf{Real Share\\ Rate}} & \multicolumn{1}{p{1cm}}{\centering \textbf{Real\\ Change}} & \multicolumn{1}{p{2.2cm}}{\centering \textbf{Additional Hypothetical Shares}} & \multicolumn{1}{p{2cm}}{\centering \textbf{Hypothetical\\ Share Rate}} & \multicolumn{1}{p{1.5cm}}{\centering \textbf{Hypothetical\\ Change}} \\ \hline
Control & \(n_0\) & \(s_0\) & \(X_0 = \frac{s_0}{n_0}\) & - & - & - & - \\
\$0.05 & \(n_1\) & \(s_1\) & \(X_1 = \frac{s_1}{n_1}\) & \(X_1 - X_0\) & \(h_1\) & \(Y_1 = \frac{s_0 + h_1}{n_0}\) & \(Y_1 - X_0 = \frac{h_1}{n_0}\) \\
\$0.20 & \(n_2\) & \(s_2\) & \(X_2 = \frac{s_2}{n_2}\) & \(X_2 - X_1\) & \(h_2\) & \(Y_2 = \frac{s_1 + h_2}{n_1}\) & \(Y_2 - X_1 = \frac{h_2}{n_1}\) \\
\$0.50 & \(n_3\) & \(s_3\) & \(X_3 = \frac{s_3}{n_3}\) & \(X_3 - X_2\) & \(h_3\) & \(Y_3 = \frac{s_2 + h_3}{n_2}\) & \(Y_3 - X_2 = \frac{h_3}{n_2}\) \\
\$1.00 & - & - & - & - & \(h_4\) & \(Y_4 = \frac{s_3 + h_4}{n_3}\) & \(Y_4 - X_3 = \frac{h_4}{n_3}\) \\ \hline
\end{tabular}
\caption{Details on methodology for comparing real and hypothetical share incentives.}
\label{tab:real_hypo_method_details}
\end{table}

Details on Table \ref{tab:real_hypo_method_details}:
\begin{itemize}
\item \(s_i =\) number of people offered \(incentive_i\) who did consent to share
\item \(h_i =\) number of people offered \(incentive_{i - 1}\) who did not consent to share but answered they would hypothetically share for \(incentive_i\)
\item \(Y_i = (s_{i-1} + h_i) / n_{i - 1}\)
\item e.g., \(Y_1 =\) (participants who either shared in control or hypothetically would share for \$0.05) / (total participants in control)
\item As a result, hypothetical change is \(h_i/n_{i-1}\) (can also be computed as \(Y_i - X_{i-1}\)).

\end{itemize}

\clearpage
\subsection{Write-in amounts for hypothetical sharing}
\label{appendix:write_in_amounts}

\begin{figure}
    \centering
    \includegraphics[width=0.65\textwidth]{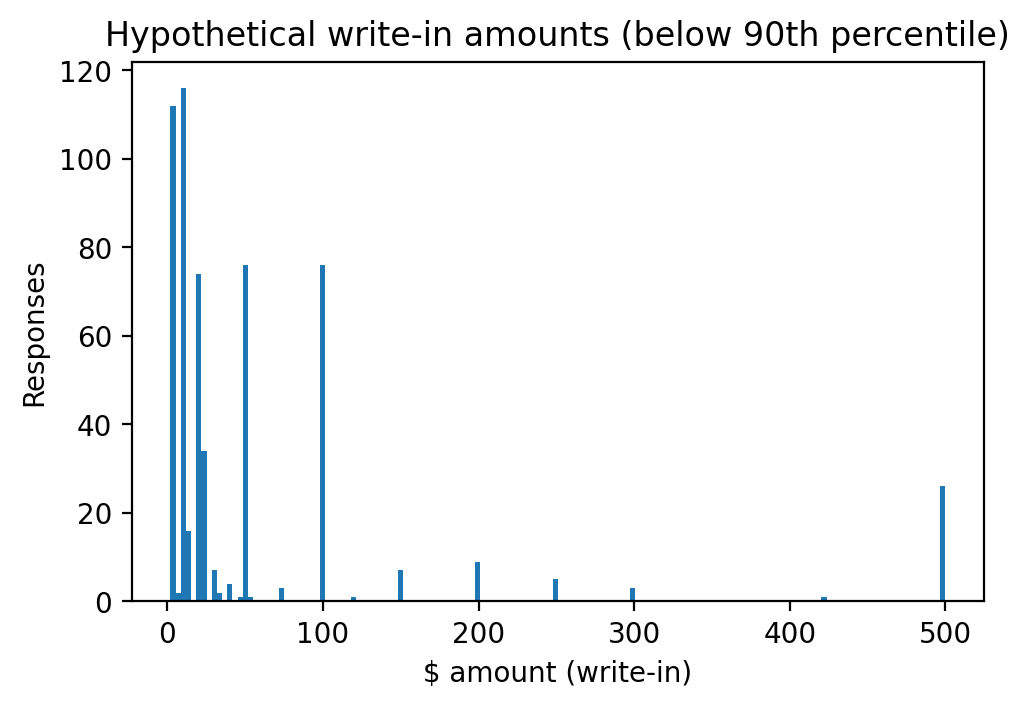}
    \caption{Histogram showing distribution of write-in values in response to what amount of money participants would hypothetically share data for, limited to values below the 90th percentile. Participants wrote in values after indicating they would not share for \$1 or less.}
    \label{fig:hypothetical_write_in_90_pct}
\end{figure}

Our experiment asked participants who did not consent to share their data whether they would hypothetically share for a (larger) bonus incentive.
Participants were asked about incrementally larger amounts up to \$1 until they said yes.
If none of these amounts were sufficient, participants were asked to specify an amount for which they would hypothetically consent to share their data or to indicate that they would not do so for any amount (see Section~\ref{sec:survey_design} for details).
Of the N=6325 participants, 641 wrote in amounts. These write-in values are not used in the main analysis. The write-in values ranged from \$2.00 to \$1,000,000,000,000,000 with a median value of \$24. 
The distribution of write-in amount values is further described by Table~\ref{tab:write_in_amounts}. Figure~\ref{fig:hypothetical_write_in_90_pct} shows a histogram depicting the distribution of write-in values below the 90th percentile.

\begin{table}[]
\centering
\small
\begin{tabular}{lc}
\hline
count & 641 \\
mean & 1.560064e+12 \\
std & 3.949763e+13\\
\hline
min & 2.00 \\
10\% & 5.00 \\
20\% & 10.00 \\
30\% & 10.00 \\
40\% & 20.00 \\
50\% & 24.00 \\
60\% & 50.00 \\
70\% &  100.00 \\
80\% & 100.00 \\
90\% & 594.00 \\
\hline
max & 1.000000e+15 \\
\hline
\end{tabular}
\caption{Distribution of values for write-in amounts.}
\label{tab:write_in_amounts}
\end{table}

\clearpage
\subsection{Data use opinions robustness check}
\label{appendix:data_use_opinions_robustness_checks}

The data use opinion questions come after participants are asked to share and presented with experimental treatments.

\textbf{Question}: 
Are responses to the data use opinion questions influenced by the incentive participants are presented? Or an interaction between this treatment and their sharing behavior?
E.g. if a participant was presented with the ‘altruism’ incentive and then opted to share, they may be more likely to answer ‘Yes’ to the question about whether researchers should be able to use consumer data.

To help answer this we add to Equation~\ref{eq:views_on_data_use_regression}, including another set of terms for incentive and share~$\times$~incentive

\begin{table}[]
\centering
\small
\begin{tabular}{p{0.38\linewidth}ccc}
\hline
\multicolumn{4}{l}{\textbf{Q1: Do you think Amazon should be able to sell YOUR purchase data to other companies?}} \\
\hline
\textbf{Predictor} & \textbf{B(log Odds)} & \textbf{Odds Ratio} & \textbf{95\% CI for Odds Ratio} \\ 
\hline
Intercept & -0.797*** & 0.451 & [0.330, 0.615] \\
share & 1.078*** & 2.939 & [2.170, 3.980] \\
\multicolumn{4}{l}{\textbf{Incentive (Reference: control)}} \\
altruism & 0.236 & 1.266 & [0.869, 1.843] \\
\$0.05 & -0.138 & 0.871 & [0.597, 1.272] \\
\$0.20 & 0.192 & 1.212 & [0.825, 1.780] \\
\$0.50 & -0.249 & 0.779 & [0.497, 1.223] \\
\multicolumn{4}{l}{\textbf{Incentive (Reference: control) $\times$ share}} \\
altruism $\times$ share & -0.227 & 0.797 & [0.523, 1.214] \\
\$0.05 $\times$ share & 0.259 & 1.296 & [0.848, 1.980] \\
\$0.20 $\times$ share & -0.23 & 0.795 & [0.518, 1.220] \\
\$0.50 $\times$ share & 0.448 & 1.565 & [0.960, 2.549] \\
\multicolumn{4}{l}{\textbf{Gender (Reference: Male)}} \\
Female & -0.647*** & 0.524 & [0.469, 0.585] \\
\multicolumn{4}{l}{\textbf{Age (Reference: 35 - 54 years)}} \\
18 - 34 years & 0.427*** & 1.533 & [1.363, 1.724] \\
55 and older & -0.554*** & 0.575 & [0.475, 0.696] \\
\multicolumn{4}{l}{\textbf{Household income (Reference: \$50,000-\$99,999)}} \\
Less than \$50,000 & 0.115 & 1.122 & [0.985, 1.277] \\
\$100,000 or more & 0.02 & 1.02 & [0.887, 1.174] \\
\multicolumn{4}{l}{\textbf{Education (Reference: Bachelor's degree)}} \\
High school or GED or less & 0.18** & 1.197 & [1.057, 1.355] \\
Graduate or professional degree & 0.103 & 1.108 & [0.951, 1.292] \\
\multicolumn{4}{l}{\textbf{Race (Reference: White)}} \\
Asian & 0.164 & 1.179 & [0.964, 1.441] \\
Black & 0.173 & 1.188 & [0.960, 1.471] \\
Other or mixed & 0.196 & 1.217 & [0.988, 1.498] \\
\multicolumn{4}{l}{\textbf{Online purchase frequency (Reference: 5-10 times/month)}} \\
Less than 5 times per month & -0.092 & 0.912 & [0.805, 1.033] \\
More than 10 times per month & 0.048 & 1.049 & [0.846, 1.302] \\ \hline
N=5820 & & &  \\
pseudo R-squared = 0.066 & & & \\ 
\hline
\end{tabular}
\caption{Robustness check for Question 1}
\label{tab:data_use_regression_Q1_robustness_check}
\end{table}

\begin{table}[]
\centering
\small
\begin{tabular}{p{0.42\linewidth}ccc}
\hline
\multicolumn{4}{p{\linewidth}}{\textbf{Q2: Do you think companies should be able to sell consumer purchase data to other companies?}} \\
\hline
\textbf{Predictor} & \textbf{B(log Odds)} & \textbf{Odds Ratio} & \textbf{95\% CI for Odds Ratio} \\ 
\hline
Intercept & -0.675*** & 0.509 & [0.374, 0.693] \\
share & 0.928*** & 2.529 & [1.872, 3.418] \\
\multicolumn{4}{p{\linewidth}}{\textbf{Incentive (Reference: control)}} \\
altruism & 0.071 & 1.073 & [0.738, 1.560] \\
\$0.05 & -0.186 & 0.83 & [0.572, 1.205] \\
\$0.20 & -0.007 & 0.993 & [0.677, 1.458] \\
\$0.50 & -0.301 & 0.74 & [0.475, 1.152] \\
\multicolumn{4}{p{\linewidth}}{\textbf{Incentive (Reference: control) $\times$ share}} \\
altruism $\times$ share & -0.057 & 0.945 & [0.621, 1.437] \\
\$0.05 $\times$ share & 0.215 & 1.24 & [0.815, 1.887] \\
\$0.20 $\times$ share & -0.041 & 0.96 & [0.626, 1.473] \\
\$0.50 $\times$ share & 0.461 & 1.586 & [0.979, 2.567] \\
\multicolumn{4}{p{\linewidth}}{\textbf{Gender (Reference: Male)}} \\
Female & -0.65*** & 0.522 & [0.468, 0.583] \\
\multicolumn{4}{p{\linewidth}}{\textbf{Age (Reference: 35 - 54 years)}} \\
18 - 34 years & 0.359*** & 1.431 & [1.273, 1.610] \\
55 and older & -0.552*** & 0.576 & [0.475, 0.698] \\
\multicolumn{4}{p{\linewidth}}{\textbf{Household income (Reference: \$50,000-\$99,999)}} \\
Less than \$50,000 & 0.076 & 1.079 & [0.948, 1.229] \\
\$100,000 or more & 0.096 & 1.101 & [0.956, 1.268] \\
\multicolumn{4}{p{\linewidth}}{\textbf{Education (Reference: Bachelor's degree)}} \\
High school or GED or less & 0.203** & 1.224 & [1.081, 1.387] \\
Graduate or professional degree & 0.075 & 1.078 & [0.924, 1.256] \\
\multicolumn{4}{p{\linewidth}}{\textbf{Race (Reference: White)}} \\
Asian & 0.145 & 1.156 & [0.944, 1.414] \\
Black & 0.484*** & 1.622 & [1.303, 2.019] \\
Other or mixed & 0.202 & 1.224 & [0.996, 1.505] \\
\multicolumn{4}{p{\linewidth}}{\textbf{Online purchase frequency (Reference: 5-10 times/month)}} \\
Less than 5 times per month & -0.025 & 0.975 & [0.861, 1.105] \\
More than 10 times per month & 0.079 & 1.082 & [0.871, 1.343] \\ \hline
\multicolumn{4}{l}{N=5750} \\
\multicolumn{4}{l}{pseudo R-squared = 0.061} \\ \hline
\end{tabular}
\caption{Robustness check for Question 2}
\label{tab:data_use_regression_Q2_robustness_check}
\end{table}

\begin{table}[]
\centering
\small
\begin{tabular}{p{0.42\linewidth}ccc}
\hline
\multicolumn{4}{p{\linewidth}}{\textbf{Q3: Big companies currently collect and sell consumer purchase data. Do you think that small businesses should be able to access this data for free in order to help them compete with the big companies?}} \\
\hline
\textbf{Predictor} & \textbf{B(log Odds)} & \textbf{Odds Ratio} & \textbf{95\% CI for Odds Ratio} \\ 
\hline
Intercept & -1.422*** & 0.241 & [0.165, 0.354] \\
share & 1.388*** & 4.007 & [2.761, 5.817] \\
\multicolumn{4}{l}{\textbf{Incentive (Reference: control)}} \\
altruism & 0.274 & 1.315 & [0.823, 2.103] \\
\$0.05 & 0.091 & 1.095 & [0.687, 1.746] \\
\$0.20 & 0.204 & 1.226 & [0.761, 1.978] \\
\$0.50 & -0.311 & 0.732 & [0.398, 1.348] \\
\multicolumn{4}{l}{\textbf{Incentive (Reference: control) $\times$ share}} \\
altruism $\times$ share & -0.054 & 0.947 & [0.567, 1.584] \\
\$0.05 $\times$ share & -0.194 & 0.823 & [0.493, 1.375] \\
\$0.20 $\times$ share & -0.49 & 0.613 & [0.364, 1.032] \\
\$0.50 $\times$ share & 0.2 & 1.221 & [0.642, 2.325] \\
\multicolumn{4}{l}{\textbf{Gender (Reference: Male)}} \\
Female & 0.133* & 1.142 & [1.009, 1.292] \\
\multicolumn{4}{l}{\textbf{Age (Reference: 35 - 54 years)}} \\
18 - 34 years & 0.258*** & 1.294 & [1.133, 1.478] \\
55 and older & -0.134 & 0.875 & [0.706, 1.083] \\
\multicolumn{4}{l}{\textbf{Household income (Reference: \$50,000-\$99,999)}} \\
Less than \$50,000 & -0.042 & 0.959 & [0.828, 1.110] \\
\$100,000 or more & -0.15 & 0.861 & [0.734, 1.009] \\
\multicolumn{4}{l}{\textbf{Education (Reference: Bachelor's degree)}} \\
High school or GED or less & 0.156* & 1.168 & [1.016, 1.344] \\
Graduate or professional degree & -0.06 & 0.942 & [0.790, 1.123] \\
\multicolumn{4}{l}{\textbf{Race (Reference: White)}} \\
Asian & 0.049 & 1.05 & [0.835, 1.321] \\
Black & -0.01 & 0.99 & [0.777, 1.260] \\
Other or mixed & 0.139 & 1.149 & [0.914, 1.445] \\
\multicolumn{4}{l}{\textbf{Online purchase frequency (Reference: 5-10 times/month)}} \\
Less than 5 times per month & -0.212** & 0.809 & [0.703, 0.930] \\
More than 10 times per month & -0.005 & 0.995 & [0.783, 1.264] \\ \hline
\multicolumn{4}{l}{N=4525} \\
\multicolumn{4}{l}{pseudo R-squared = 0.052} \\ \hline
\end{tabular}
\caption{Robustness check for Question 3}
\label{tab:data_use_regression_Q3_robustness_check}
\end{table}

\begin{table}[]
\centering
\small
\begin{tabular}{p{0.42\linewidth}ccc}
\hline
\multicolumn{4}{p{\linewidth}}{\textbf{Q4: Do you think the U.S. Census Bureau should use purchase data to supplement their existing surveys?}} \\
\hline
\textbf{Predictor} & \textbf{B(log Odds)} & \textbf{Odds Ratio} & \textbf{95\% CI for Odds Ratio} \\ 
\hline
Intercept & -1.767*** & 0.171 & [0.113, 0.257] \\
share & 1.061*** & 2.89 & [1.940, 4.306] \\
\multicolumn{4}{l}{\textbf{Incentive (Reference: control)}} \\
altruism & 0.005 & 1.005 & [0.602, 1.679] \\
\$0.05 & -0.175 & 0.839 & [0.502, 1.404] \\
\$0.20 & 0.048 & 1.049 & [0.624, 1.764] \\
\$0.50 & 0.237 & 1.268 & [0.715, 2.248] \\
\multicolumn{4}{l}{\textbf{Incentive (Reference: control) $\times$ share}} \\
altruism $\times$ share & 0.06 & 1.062 & [0.607, 1.860] \\
\$0.05 $\times$ share & 0.131 & 1.14 & [0.648, 2.003] \\
\$0.20 $\times$ share & -0.053 & 0.949 & [0.538, 1.672] \\
\$0.50 $\times$ share & -0.334 & 0.716 & [0.387, 1.324] \\
\multicolumn{4}{l}{\textbf{Gender (Reference: Male)}} \\
Female & -0.56*** & 0.571 & [0.499, 0.653] \\
\multicolumn{4}{l}{\textbf{Age (Reference: 35 - 54 years)}} \\
18 - 34 years & 0.549*** & 1.732 & [1.500, 1.999] \\
55 and older & -0.331** & 0.718 & [0.561, 0.919] \\
\multicolumn{4}{l}{\textbf{Household income (Reference: \$50,000-\$99,999)}} \\
Less than \$50,000 & -0.001 & 0.999 & [0.851, 1.173] \\
\$100,000 or more & 0.21* & 1.234 & [1.043, 1.460] \\
\multicolumn{4}{l}{\textbf{Education (Reference: Bachelor's degree)}} \\
High school or GED or less & -0.003 & 0.997 & [0.856, 1.162] \\
Graduate or professional degree & 0.245** & 1.278 & [1.063, 1.536] \\
\multicolumn{4}{l}{\textbf{Race (Reference: White)}} \\
Asian & 0.518*** & 1.679 & [1.332, 2.116] \\
Black & 0.27* & 1.31 & [1.016, 1.690] \\
Other or mixed & 0.073 & 1.076 & [0.838, 1.383] \\
\multicolumn{4}{l}{\textbf{Online purchase frequency (Reference: 5-10 times/month)}} \\
Less than 5 times per month & -0.019 & 0.981 & [0.842, 1.143] \\
More than 10 times per month & 0.169 & 1.184 & [0.916, 1.529] \\ \hline
\multicolumn{4}{l}{N=4390} \\
\multicolumn{4}{l}{pseudo R-squared = 0.060} \\ \hline
\end{tabular}
\caption{Robustness check for Question 4}
\label{tab:data_use_regression_Q4_robustness_check}
\end{table}

\begin{table}[]
\centering
\small
\begin{tabular}{p{0.42\linewidth}ccc}
\hline
\multicolumn{4}{p{\linewidth}}{\textbf{Q5: Do you think researchers should be able to use purchase data to understand societal changes (e.g. due to COVID-19)?}} \\
\hline
\textbf{Predictor} & \textbf{B(log Odds)} & \textbf{Odds Ratio} & \textbf{95\% CI for Odds Ratio} \\ 
\hline
Intercept & -0.381* & 0.683 & [0.488, 0.957] \\
share & 1.34*** & 3.82 & [2.752, 5.302] \\
\multicolumn{4}{p{\linewidth}}{\textbf{Incentive (Reference: control)}} \\
altruism & 0.086 & 1.089 & [0.729, 1.627] \\
0.05 & 0.087 & 1.091 & [0.742, 1.604] \\
0.20 & 0.093 & 1.098 & [0.731, 1.649] \\
0.50 & 0.055 & 1.057 & [0.665, 1.678] \\
\multicolumn{4}{p{\linewidth}}{\textbf{Incentive (Reference: control) $\times$ share}} \\
altruism $\times$ share & 0.179 & 1.196 & [0.749, 1.909] \\
0.05 $\times$ share & 0.001 & 1.001 & [0.636, 1.575] \\
0.20 $\times$ share & -0.12 & 0.887 & [0.555, 1.417] \\
0.50 $\times$ share & -0.032 & 0.969 & [0.578, 1.624] \\
\multicolumn{4}{p{\linewidth}}{\textbf{Gender (Reference: Male)}} \\
Female & -0.057 & 0.945 & [0.828, 1.077] \\
\multicolumn{4}{p{\linewidth}}{\textbf{Age (Reference: 35 - 54 years)}} \\
18 - 34 years & 0.292*** & 1.339 & [1.162, 1.543] \\
55 and older & -0.252* & 0.778 & [0.627, 0.964] \\
\multicolumn{4}{p{\linewidth}}{\textbf{Household income (Reference: \$50,000-\$99,999)}} \\
Less than \$50,000 & 0.123 & 1.131 & [0.968, 1.321] \\
\$100,000 or more & 0.174* & 1.191 & [1.005, 1.411] \\
\multicolumn{4}{p{\linewidth}}{\textbf{Education (Reference: Bachelor's degree)}} \\
High school or GED or less & -0.083 & 0.921 & [0.795, 1.067] \\
Graduate or professional degree & 0.293** & 1.341 & [1.107, 1.624] \\
\multicolumn{4}{p{\linewidth}}{\textbf{Race (Reference: White)}} \\
Asian & 0.225 & 1.252 & [0.973, 1.611] \\
Black & -0.149 & 0.861 & [0.672, 1.104] \\
Other or mixed & 0.003 & 1.003 & [0.783, 1.284] \\
\multicolumn{4}{p{\linewidth}}{\textbf{Online purchase frequency (Reference: 5-10 times/month)}} \\
Less than 5 times per month & 0 & 1 & [0.861, 1.160] \\
More than 10 times per month & -0.03 & 0.97 & [0.752, 1.251] \\ \hline
\multicolumn{4}{l}{N=4910} \\
\multicolumn{4}{l}{pseudo R-squared = 0.060} \\ \hline
\end{tabular}
\caption{Robustness check for Question 5}
\label{tab:data_use_regression_Q5_robustness_check}
\end{table}

\clearpage
\section{Amazon data}
\label{SI:amazon_data}

This section includes descriptive statistics for the Amazon purchases dataset that was collected through the data collection tool that is described in this paper. It also includes example analyses to demonstrate the dataset's effectiveness for future research. Readers are encouraged to access the openly published dataset for use in their own analyses.

\begin{table}[h]
\begin{adjustwidth}{-2cm}{-2cm} 
\centering
\tiny
\begin{tabular}
{p{1cm}p{0.9cm}p{0.7cm}p{0.8cm}p{2.5cm}llp{2cm}}
\hline
\textbf{Order Date} & \textbf{Purchase Price Per Unit} & \textbf{Quantity} & \textbf{Shipping Address State} & \textbf{Title} & \textbf{ASIN/ISBN} & \textbf{Category} & \textbf{Survey ResponseID} \\ \hline
2018-01-21 & \$23.07 & 1.0 & OK & OTTERBOX SYMMETRY SERIES Case for iPhone 8 PLUS \& iPhone 7 PLUS (ONLY) - Frustration Free Packaging - SALTWATER TAFFY (PIPELINE PINK/BLAZER BLUE) & B01K6PBRSW & CELLULAR\_PHONE\_CASE & R\_2zARigFdY655hAS \\ \hline
2018-02-06 & \$15.91 & 1.0 & OK & Strength in Stillness: The Power of Transcendental Meditation & 1501161210 & ABIS\_BOOK & R\_2zARigFdY655hAS \\ \hline
2018-04-03 & \$5.99 & 1.0 & OK & Square Reader for magstripe (with headset jack) & B00HZYK3CO & MEMORY\_CARD\_READER & R\_2zARigFdY655hAS \\ \hline
2018-06-11 & \$4.89 & 1.0 & OK & Dove Advanced Care Antiperspirant Deodorant Stick for Women, Original Clean, for 48 Hour Protection And Soft And Comfortable Underarms, 2.6 oz & B00Q70R41U & BODY\_DEODORANT & R\_2zARigFdY655hAS \\ \hline
\end{tabular}
\caption{A representative sample of rows from one respondent's Amazon data.}
\label{tab:sample_data}
\end{adjustwidth}
\end{table}

The Amazon purchase histories dataset includes a total of 1,850,717 purchases from 5027 respondents who shared their data.
Table~\ref{tab:sample_data} shows a sample of rows from the Amazon data from one randomly selected respondent.

\begin{table}[h]
\centering
\small
\begin{tabular}{p{6cm}P{2cm}P{1.5cm}P{1.5cm}}
\hline
\textbf{Product Title} & \textbf{Distinct Users Making Purchases} & \textbf{Total Purchases} & \textbf{Total Spend} \\ 
\hline
Echo Dot (3rd Gen, 2018 release) - Smart speaker with Alexa - Charcoal & 377 & 484 & \$13,195.60 \\
\hline
Amazon Basics 36 Pack AAA High-Performance Alkaline Batteries, 10-Year Shelf Life, Easy to Open Value Pack & 366 & 571 & \$6,321.26 \\
\hline
Fire TV Stick 4K streaming device with Alexa Voice Remote (includes TV controls) | Dolby Vision & 350 & 461 & \$20,670.05 \\
\hline
Amazon Basics 48 Pack AA High-Performance Alkaline Batteries, 10-Year Shelf Life, Easy to Open Value Pack & 305 & 576 & \$8,641.42 \\
\hline
Amazon Smart Plug, works with Alexa – A Certified for Humans Device & 290 & 353 & \$8,428.02 \\ \hline
\end{tabular}
\caption{Top 5 products, number of distinct users purchasing the product, total purchases, and total spend, sorted by number of purchasing users, when excluding gift cards.}
\label{tab:top_products}
\end{table}

Table~\ref{tab:top_products} shows the top 5 products by their Title, when sorting by the number of distinct users making purchases for the corresponding ASIN/ISBN (Product Code), and when excluding gift cards.

\begin{table}[h]
\centering
\small
\begin{tabular}{lP{2cm}P{1.5cm}P{1.5cm}}
\hline
\textbf{Item Category} & \textbf{Distinct Users Making Purchases} & \textbf{Total Purchases} & \textbf{Total Spend} \\
\hline
ABIS\_BOOK & 4236 & 87,619 & \$1,359,183.61 \\
ELECTRONIC\_CABLE & 3521 & 18,268 & \$222,390.71 \\
CELLULAR\_PHONE\_CASE & 3468 & 15,370 & \$229,662.82 \\
SHIRT & 3365 & 27,267 & \$514,584.54 \\
HEADPHONES & 3307 & 11,394 & \$546,323.79 \\
\hline
\end{tabular}
\caption{Top 5 product categories, number of distinct users purchasing products in the category, total purchases, and total spend, sorted by number of users. Note there are product codes where the category changes over time in the longitudinal data.}
\label{tab:top_categories}
\end{table}

Table~\ref{tab:top_categories} shows data for the top 5 product categories when aggregating purchases by the ``Category`` column and sorting by the number of distinct users making the purchases. 
The tables also report on the total number of purchases and total spend for these categories. 

\begin{figure}[h]
    \centering
    \includegraphics[width=\textwidth]{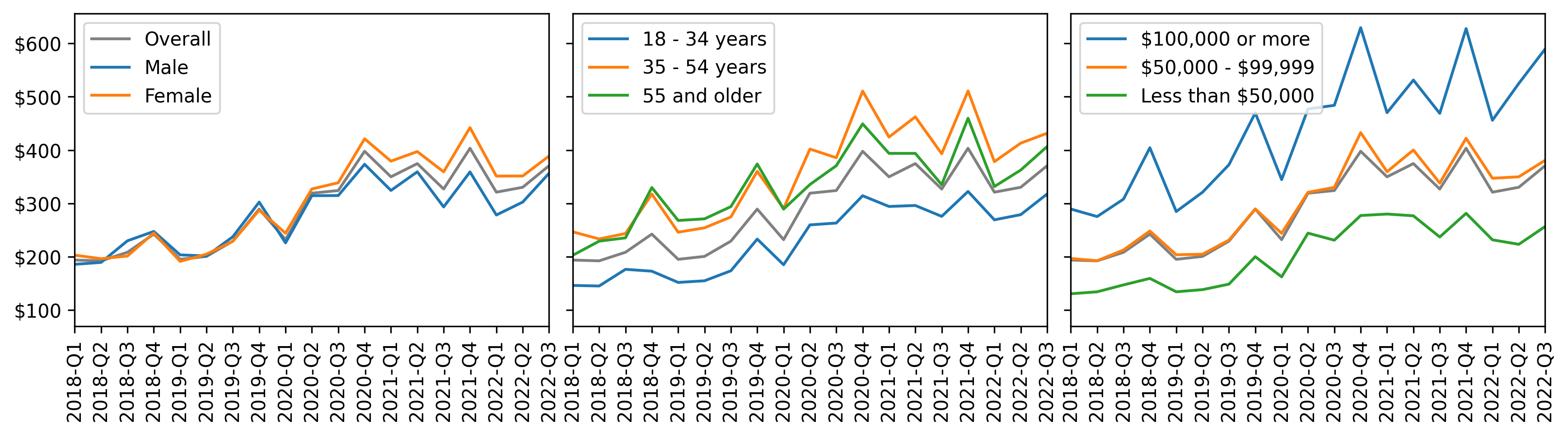}
    \caption{Quarterly median user spend by demographic group, compared to median user spend overall (gray). Left: Spend for Male vs Female users. Middle: Spend by age. Right: Spend by household income. }
    \label{fig:med_spend_by_demographic}
\end{figure}

Figure~\ref{fig:med_spend_by_demographic} shows time series plots for the median spend per user, for each quarter, and highlights differences across demographic groups. 
A gray line shows the median user spend overall. 
While the demographic groups in Figure~\ref{fig:med_spend_by_demographic} are limited to the Male/Female binary and users who provided their household income, all users, including those who answered ``Other`` or ``Prefer not to say``, are included in the calculation of overall median spend. 
The left plot shows the difference between Male and Female users.
The middle plot shows differences between age groups. The right plot shows differences by household income. As might be expected, users with higher incomes spend more on average, especially in the Q4 holiday season.
There are also notable differences in spending by age group, where younger users spend less on average, as well as by gender, where female users spend more on average after the start of COVID-19 (2020-Q2).

With the above differences in purchasing behaviors and sampling biases in mind, we use stratified random sampling, without replacement, to create a stratified sample of users. 
The strata are defined by both age and sex and match population proportions reported in 2022 U.S. Census data~\cite{NC-EST2022-AGESEX-RES}.
In particular, strata are defined by a binary definition of sex (Male, Female) and age groups aggregated to 3 levels (18-34, 35-54, 55 and older), as shown in Figure~\ref{fig:med_spend_by_demographic}, resulting in 6 strata. 

\begin{figure}[h]
    \centering
    \includegraphics[width=\textwidth]{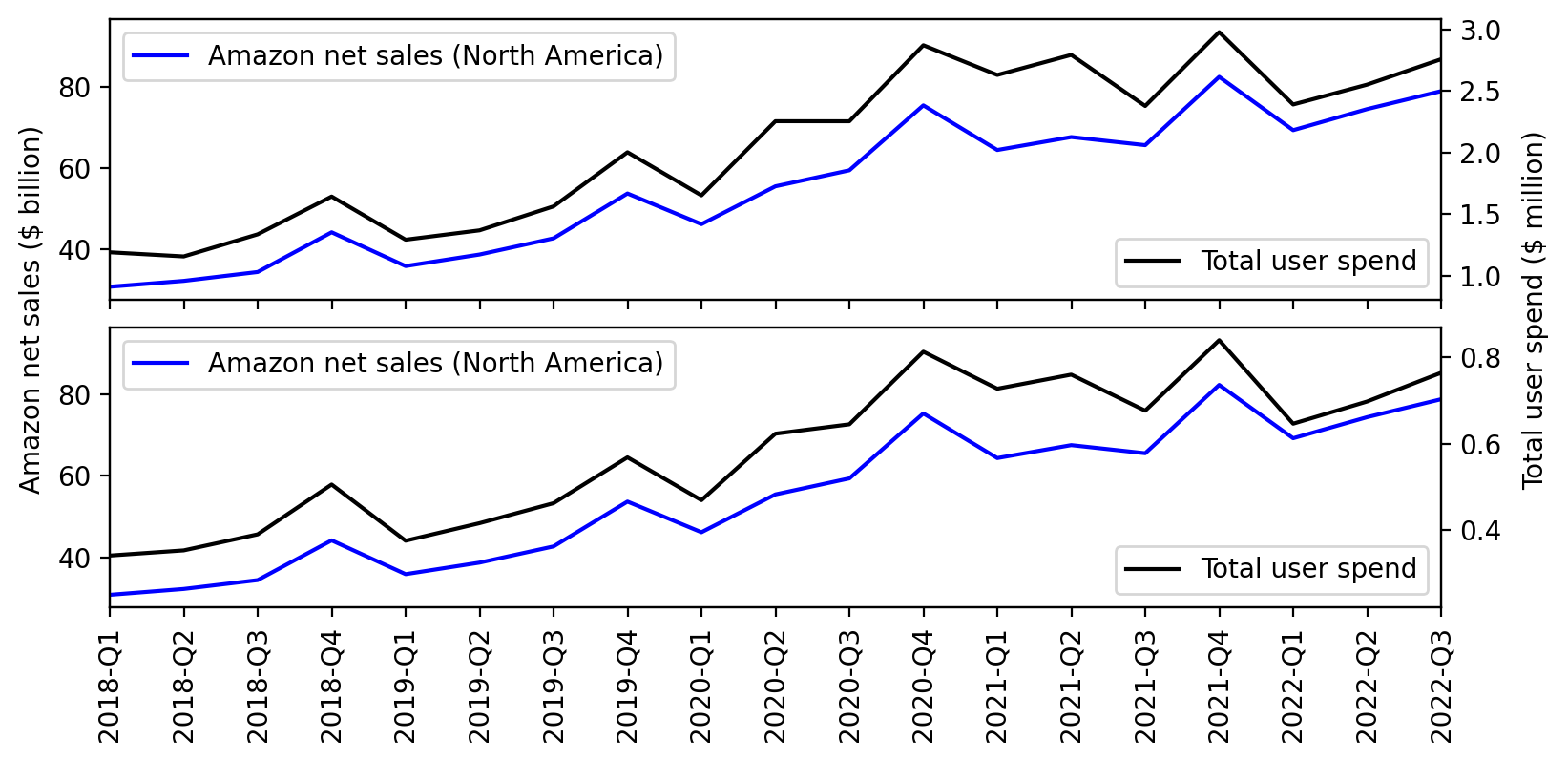}
    \caption{Quarterly Amazon net sales (North America segment) and total user spend. Data are highly correlated (p$<.001$). Top: Data are shown for total user spend for the entire sample. Bottom: Data are shown for total user spend for the stratified sample.}
    \label{fig:quarterly_spend_amzn_users}
\end{figure}

In order to assess how representative our dataset is for Amazon purchasing in general, we compare Amazon net sales data (for the North America segment) to total spend by users in our sample, for each quarter in our studied period. 
Figure~\ref{fig:quarterly_spend_amzn_users} shows this comparison. The top plot compares Amazon sales data to total user spend for our full sample while the bottom plot restricts the total user spend data to the stratified sample.
Amazon quarterly net sales data are from their quarterly earnings releases produced for investor relations\footnote{Amazon's quarterly earnings releases are available at \url{https://ir.aboutamazon.com/quarterly-results/default.aspx}.}.
There are important differences in these sales data sources that we compare:
The Amazon net sales data include all of North America, while our purchases dataset is limited to the U.S. Furthermore, our data is for a consistent sample of users with accounts starting in 2018 and does not account for increased sales due to new Amazon users.
Despite these differences, the quarterly Amazon sales data and total user spend are highly correlated. The Pearson coefficient is r=.978 (p<.001) with data from the entire sample and r=.975 (p<.001) with data from the stratified sample.

\begin{figure}
    \centering
    \includegraphics[width=\textwidth]{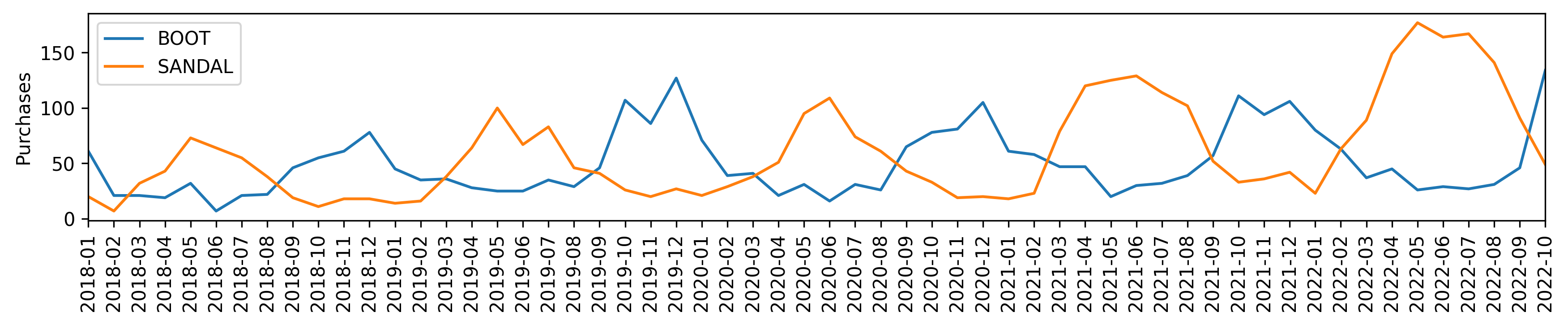}
    \caption{Total purchases each month for categories BOOT and SANDAL. Purchases for these products demonstrate different seasonal trends present in the dataset, where SANDAL purchases have yearly peaks in the summer months while BOOT purchases have yearly peaks in the winter months.}
    \label{fig:seasonal_footwear_purchases}
\end{figure}

We also show the potential utility of the Amazon purchases data when considering specific product types and seasonal trends. 
Figure~\ref{fig:seasonal_footwear_purchases} plots the total monthly purchases for products in the dataset with category ``BOOT`` and products with the category ``SANDAL``. 
Total purchases are computed by summing over the quantity in each such purchase row.
As to be expected, purchases for these products demonstrate opposite seasonality trends, where SANDAL purchases have yearly peaks in the summer months while BOOT purchases have yearly peaks in the winter months.

\begin{figure}
    \centering
    \includegraphics[width=\textwidth]{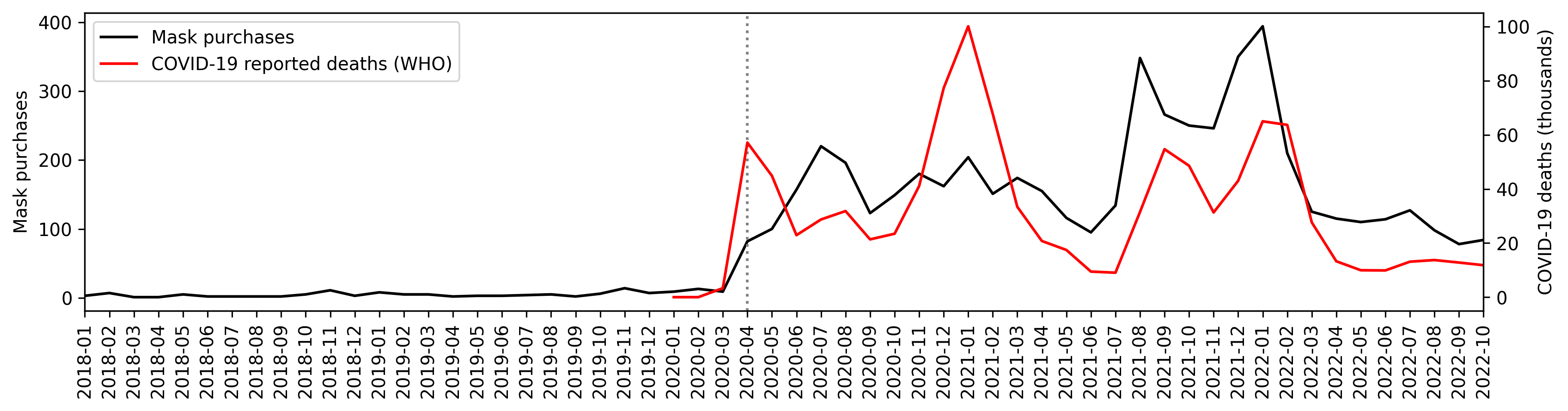}
    \caption{Monthly COVID-19 reported deaths (U.S. data reported by WHO) compared to face mask purchases.}
    \label{fig:COVID_19_purchases_deaths}
\end{figure}

Figure~\ref{fig:COVID_19_purchases_deaths} shows a timeseries of the monthly reported COVID-19 deaths in the entire U.S. compared to total number of face mask purchases in our dataset.
The COVID-19 data are from the World Health Organization (WHO)~\cite{WHO_COVID19}. 
Figure~\ref{fig:COVID_19_purchases_deaths} shows how both the face mask purchases and COVID-19 deaths have a clear initial spike at the start of the COVID-19 pandemic in April 2020.
These metrics continue to have similar trends, with spikes in the winter months and when students began to return to school in August and September 2021.

We look forward to the future analyses that may be produced using this open dataset by either ourselves or other researchers.

\end{document}